\documentstyle[seceq,preprint]{ptptex}

\newcommand{\bra}[1]{\langle {#1} |}     
\newcommand{\ket}[1]{| {#1} \rangle}     
\newcommand{\bbra}[1]{\langle\langle {#1} |}     
\newcommand{\kket}[1]{| {#1} \rangle\rangle}     
\newcommand{\rket}[1]{| {#1} )}     
\newcommand{\dket}[1]{|| {#1} \rangle}     
\newcommand{\maru}[1]{\widetilde {#1}} 
\newcommand{\wtilde}[1]{{\widetilde #1}}     

\title{
A Possible Boson Realization of Generalized Lipkin Model 
for Many-Fermion System
}
\subtitle{The su(M+1)-Algebraic Model\\
in Non-Symmetric 
Boson Representation}    

\author{
Atsushi {\sc Kuriyama}, Constan\c{c}a {\sc Provid\^encia}${}^*$, 
Jo\~ao da {\sc Provid\^encia}$^{*}$, \\
Yasuhiko {\sc Tsue}${}^{**}$ and Masatoshi {\sc Yamamura}
}

\inst{
Faculty of Engineering, Kansai University, Suita 564-8680, Japan\\
$^{*}$Departamento de Fisica, Universidade de Coimbra, P-3000 Coimbra, 
Portugal\\
$^{**}$Department of Material Science, Kochi University, Kochi 780-8520, 
Japan
}

\recdate{
}

\abst{
On the basis of the formalism proposed by three of the present 
authors (A. K., J. P. \& M. Y.), generalized Lipkin model 
consisting of $(M+1)$ single-particle levels is investigated. 
This model is essentially a kind of the $su(M+1)$-algebraic model 
and, in contrast to the conventional treatment, the case, where 
fermions are partially occupied in each level, is discussed. 
The scheme for obtaining the orthogonal set for the irreducible 
representation is presented. 
}

\begin{document}

\maketitle

\section{Introduction and preliminaries}

The Lipkin model,\cite{Lipkin}
which was proposed by Lipkin, Meshkov and Glick in 1965, has 
played a peculiar role in schematic studies of collective 
motions in many-fermion system. 
This is a kind of shell model : 
Under certain interaction, many fermions move in two 
single-particle levels with the same degeneracy. 
This is essentially one kind of the $su(2)$-algebraic model 
and it has also contributed to schematic understanding 
of finite temperature effects in many-fermion system.\cite{Brito}
On the other hand, the Lipkin model has been generalized to the case 
of many single-particle levels, for example, the case of three levels 
is the most popular and it is a kind of the $su(3)$-algebraic model. 
The investigations based on this model is also not only 
for collective motion\cite{Li} but also for finite temperature 
effects in many-fermion system.\cite{Terra}

The generators of the $su(2)$ algebra, which we denote 
$({\maru S}_+, {\maru S}_-, {\maru S}_0)$, play a central role 
in the Lipkin model. 
These are expressed in terms of the bilinear forms for 
particle operators $({\maru \alpha}_{1m}, {\maru \alpha}_{1m}^*)$ 
and hole operators 
$({\maru \beta}_{\widetilde m}, {\maru \beta}_{\widetilde m}^*)$ : 
\begin{subequations}\label{1-1}
\begin{eqnarray}\label{1-1a}
& &{\maru S}_+
=\sum_m {\maru \alpha}_{1m}^*{\maru \beta}_{\widetilde m}^* \ , 
\qquad
{\maru S}_-
=\sum_m {\maru \beta}_{\widetilde m}{\maru \alpha}_{1m} \ , 
\nonumber\\
& &{\maru S}_0=\frac{1}{2}\left[
\sum_m{\maru \alpha}_{1m}^*{\maru \alpha}_{1m}
-(\Omega-\sum_{m}{\maru \beta}_{\widetilde m}^*{\maru \beta}_{\widetilde m})
\right] \ .
\end{eqnarray}
Here, $\Omega$ denotes the degeneracy of the single-particle 
levels. The definitions of the other notations will be given 
in \S 2. 
For the later convenience, we will use 
$({\maru S}^1, {\maru S}_1, {\maru S}_1^1)$ for the 
$su(2)$ generators defined as 
\begin{equation}\label{1-1b}
{\maru {S}}^1={\maru S}_+ \ , \quad
{\maru S}_1={\maru S}_- \ , \quad
{\maru S}_1^1=2{\maru S}_0 \ .
\end{equation}
\end{subequations}
In addition to the above generators, there exists one operator, 
which we denote ${\maru N}$ as 
\begin{equation}\label{1-2}
{\maru N}=(\Omega-\sum_m{\maru \beta}_{\widetilde m}^*
{\maru \beta}_{\widetilde m})
+\sum_m{\maru \alpha}_{1m}^*{\maru \alpha}_{1m} \ .
\end{equation}
The operator ${\maru N}$ denotes the total fermion number 
and it obeys
\begin{equation}\label{1-3}
\left[ {\maru N} , {\maru S}^1 \right] =
\left[ {\maru N} , {\maru S}_1 \right] =
\left[ {\maru N} , {\maru S}_1^1 \right] = 0 \ .
\end{equation}
Conventionally, the orthogonal set for the Lipkin model 
is obtained by operating ${\maru S}^1$ successively 
on the state $\kket{m}$ which obeys the condition 
\begin{equation}\label{1-4}
{\maru S}_1\kket{m}=0 \ , \qquad
{\maru S}_1^1\kket{m}=-\sigma_1\kket{m} \ .
\end{equation}
The state $\kket{m}$ is the eigenstate of the Casimir operator 
of the $su(2)$ algebra : 
\begin{eqnarray}\label{1-5}
& &{\maru \Gamma}_{su(2)}\kket{m}
=\frac{1}{2}\sigma_1(\sigma_1+2)\kket{m} \ , \nonumber\\
& &{\maru \Gamma}_{su(2)}={\maru S}^1{\maru S}_1
+{\maru S}_1{\maru S}^1+\frac{1}{2}\left({\maru S}_1^1\right)^2 \ .
\end{eqnarray}
The quantity $\sigma_1/2$ denotes the magnitude of the quasi-spin. 
Further, the relation (\ref{1-3}) supports that $\kket{m}$ 
is also the eigenstate of ${\maru N}$ : 
\begin{equation}\label{1-6}
{\maru N}\kket{m}=N\kket{m} \ .
\end{equation}
By operating ${\maru S}^1$ for 
$(\sigma_1+\sigma_0)/2$ times on $\kket{m}$, we have 
\begin{eqnarray}\label{1-7}
& &\kket{(\gamma); N, \sigma_1, \sigma_0}
=\left({\maru S}^1\right)^{(\sigma_1+\sigma_0)/2}\kket{m} \ , \nonumber\\
& &\kket{m}=\kket{(\gamma); N, \sigma_1} \ .
\end{eqnarray}
Here, $(\gamma)$ denotes a set of the quantum numbers 
additional to those related to the $su(2)$ algebra. 
The definitions of ${\maru S}_1^1$ and ${\maru N}$ 
tell us that $\kket{m}$ can be specified by the quantum 
numbers $n$ and $n_1$ which are the eigenvalues 
of $\sum_m{\maru \beta}_{\widetilde m}^*{\maru \beta}_{\widetilde m}$ and 
$\sum_m{\maru \alpha}_{1m}^*{\maru \alpha}_{1m}$, respectively : 
\begin{equation}\label{1-8}
\kket{m}=\kket{(\gamma); n, n_1} \ .
\end{equation}
The state $\kket{m}$ is called the minimum weightly state, 
but, in this paper, we will call it the intrinsic state 
in the meaning analogous to the rotational model. 
The relation between $(N, \sigma_1)$ and $(n, n_1)$ is 
given as 
\begin{subequations}\label{1-9}
\begin{eqnarray}
& &N=\Omega-n+n_1 \ , \qquad\quad
\sigma_1=\Omega-n-n_1 \ , \label{1-9a}\\
& &n=\Omega-\frac{1}{2}(N+\sigma_1) \ , \qquad
n_1=\frac{1}{2}(N-\sigma_1) \ . \label{1-9b}
\end{eqnarray}
\end{subequations}
Since $0\le \Omega-n \le \Omega$ and 
$0\le n_1 \le \Omega$, we have the following relations :
\begin{subequations}\label{1-10}
\begin{eqnarray}
& &{\rm if}\ \ 0\le N \le \Omega \ , \qquad\quad
0\le \sigma_1 \le N \ , \label{1-10a}\\
& &{\rm if}\ \ \Omega\le N \le 2\Omega \ , \qquad\ \ 
0\le \sigma_1 \le 2\Omega-N \ , \label{1-10b}
\end{eqnarray}
\end{subequations}
The above is an outline of the Lipkin model and it is easily 
generalized to the $su(M+1)$-algebraic model in the 
$(M+1)$ single-particle levels with the same degeneracy, 
which will be discussed in \S 3.

Our present interest is concerned with the explicit determination 
of $\kket{m}$ in terms of the fermion operators. 
With the use of the quasi-fermion operators obeying 
certain constraints,\cite{YK}but, it may be difficult to 
apply this idea to the generalized case given in \S 3. 
However, one case is simply given : $n=n_1=0$, 
i.e., $N=\sigma_1=\Omega$. 
In this case, $\kket{m}$ corresponds to the vacuum of 
${\maru \alpha}_{1m}$ and ${\maru \beta}_{\widetilde m}\ 
({\maru \alpha}_{1m}\kket{m}={\maru \beta}_{\widetilde m}\kket{m}
=0)$ and it means that one level is fully occupied by 
the fermions, i.e., 
the closed shell system. 
The case of the generalized Lipkin model is in the same 
situation as that in the $su(2)$ model. 
Except the case of the finite temperature effects,\cite{Brito}
many of the investigations based on the Lipkin model 
is restricted to the case $n=n_1=0$, that is, the use 
of the orthogonal set obtained by operating 
${\maru S}^1$ on the state $\kket{(\gamma); n=0,n_1=0}\ 
(=\kket{(\gamma); N=\Omega, \sigma_1=\Omega})$. 
In Ref.\citen{Brito}, all the values of $\sigma_1$ permitted, which are shown 
in the relation (\ref{1-10a}), are taken into account for the system 
with $N=\Omega$. 
The case of the $su(3)$ Lipkin model\cite{Li} is 
also in the situation similar to the above except the 
finite temperature effects.\cite{Terra}

As is well known, there exist two forms for the boson realization 
of the Lipkin model. 
One is the Holstein-Primakoff representation\cite{HP} 
and the other the Schwinger representation.\cite{S} 
In the former, the three generators are expressed as 
\begin{eqnarray}\label{1-11}
& &{\widetilde {\cal S}}^1={\widetilde C}^*\cdot
\sqrt{S_1-{\widetilde C}^*{\widetilde C}} \ , \qquad
{\widetilde {\cal S}}_1=
\sqrt{S_1-{\widetilde C}^*{\widetilde C}}\cdot{\widetilde C} \ , 
\nonumber\\
& &{\widetilde {\cal S}}_1^1=2{\widetilde C}^*{\wtilde C}-S_1 \ .
\end{eqnarray}
Here, $({\wtilde C}, {\wtilde C}^*)$ denote boson operators and 
$\sigma_1$ is given in the relation (\ref{1-5}). 
The state $\rket{s_1, s_0}$ is constructed in the form 
\begin{eqnarray}\label{1-12}
& &\rket{s_1,s_0}=({\widetilde {\cal S}}^1)^{(s_1+s_0)/2}
\rket{m}=({\wtilde C}^*)^{(s_1+s_0)/2}\rket{0} \ , \nonumber\\
& &\rket{m}=\rket{0} \ . \quad
({\wtilde C}\rket{0}=0)
\end{eqnarray}
The Schwinger representation gives the following form for 
the generators : 
\begin{equation}\label{1-13}
{\hat S}^1={\hat a}_1^*{\hat b} \ , \qquad
{\hat S}_1={\hat b}^*{\hat a}_1 \ , \qquad
{\hat S}_1^1={\hat a}_1^*{\hat a}_1-{\hat b}^*{\hat b} \ .
\end{equation}
Here, $({\hat a}_1, {\hat a}_1^*)$ and 
$({\hat b}, {\hat b}^*)$ denote two kinds of bosons. 
The state $\ket{s_1, s_0}$ is constructed in the form 
\begin{eqnarray}\label{1-14}
& &\ket{s_1, s_0}=({\hat S}^1)^{(s_1+s_0)/2}\ket{m}
=({\hat a}_1^*)^{(s_1+s_0)/2}({\hat b}^*)^{(s_1-s_0)/2}\ket{0} \ ,
\nonumber\\
& &\ket{m}=({\hat b}^*)^{s_1}\ket{0}\ . \quad
({\hat a}_1\ket{0}={\hat b}\ket{0}=0)
\end{eqnarray}
In a certain case mentioned below, the Holstein-Primakoff 
representation is easily generalized and it is called the 
symmetric representation.\cite{KM}

The above two boson realizations have undoubtedly contributed in 
the development of the studies of many-fermion systems. 
In order to connect the above boson systems with the original 
fermion system, it may be natural to set up the 
following condition : 
\begin{equation}\label{1-15}
s_1=\sigma_1 \ (=\Omega-n-n_1=2(\Omega-n)-N) \ .
\end{equation}
The relation (\ref{1-15}) comes from the relation (\ref{1-9a}). 
Of course, the quantity $s_1$ obeys the restrictions (\ref{1-10}). 
However, usually, the above two boson realizations have been 
applied to the case where one level is fully occupied : 
$s_1=N=\Omega$, i.e., $n=n_1=0$. 
This means that, with the help of these forms, 
we are able to obtain only schematic knowledges on the low-lying 
states of so-called closed shell system. 
In other words, it may be impossible to get any information 
not only on the low-lying states of open shell systems but also on 
the high-lying states of closed or open shell systems. 
Further, as was already mentioned, the Holstein-Primakoff 
representation was generalized in the frame of the symmetric 
representation, which enable us to describe the closed 
shell systems. 
If we intend to investigate the above-mentioned cases in 
the frame of the boson space, we must generalize the 
boson realization of the Lipkin model in the form 
containing the condition (\ref{1-15}), i.e., in the 
form of non-symmetric representation. 

Main aim of this paper is to present a possible boson 
realization of generalized Lipkin model in non-symmetric representation. 
We investigate a model consisting of $(M+1)$ single-particle levels 
with the same degeneracy. 
This is nothing but generalized Lipkin model.\cite{KM} 
In the intrinsic state for this model, each single-particle 
level is partially occupied by the fermions. 
This is the most important point of the present investigation. 
For this investigation, a general framework for the $su(M+1)$ algebra 
in the Schwinger boson representation may be useful. 
This framework was proposed by three of the present 
authors (A. K., J. P. \& M. Y.), which will 
be, hereafter, referred to as (I).\cite{KPY1} 
In (I), $(M+1)(N+1)$ kinds of boson operators are prepared. 
With the use of them, we can construct $((M+1)^2-1)$ generators 
of the $su(M+1)$ algebra in terms of the bilinear forms 
for the bosons. 
On the other hand, the $su(N,1)$ algebra, which is independent 
of the above $su(M+1)$ algebra, is constructed in the 
frame of the above bosons. 
Further, it can be seen that there exists one operator, 
which commutes with all the generators of both algebras. 
This operator plays a role similar to that of the fermion 
number operator in the Lipkin model. 
With the aid of this operator, we can formulate the 
$su(M+1)$ and the $su(N,1)$ algebraic model and under certain 
correspondence between the fermion and the boson space, 
we are able to have the generalized Lipkin model in the 
framework of the Schwinger boson representation developed 
in (I). 
The conventional form is called the symmetric boson representation 
and the present one may be called non-symmetric boson representation. 
Especially, the intrinsic state with the partially occupied 
single-particle levels can be expressed explicitly 
in terms of the boson operators. 
Under the general scheme for obtaining the orthogonal set, 
we can show two cases, the $su(2)$ and the $su(3)$ algebra 
in the explicit form.

In next section, we will recapitulate the essential 
part of (I) for the present discussion. 
Section 3 will be devoted to giving the generalized Lipkin 
model in the fermion space. 
In \S 4, we will discuss the method how to construct 
the intrinsic state in the boson space. 
Through this discussion, it is enough for the present 
aim to consider the case $M=N$ for the $su(M+1)$ and the 
$su(N,1)$ algebra. 
In \S 5, the correspondence between the intrinsic states 
for the fermion and the boson space will be discussed. 
Section 6 will be devoted to giving the general scheme 
for constructing the orthogonal set for the irreducible 
representation. 
Finally, in \S 7, the cases of the $su(2)$ and the $su(3)$ 
algebras will be explicitly presented and the concluding 
remarks will be mentioned. 
In Appendices, some mathematical formulae and proof, which are 
needed in this paper, will be given.

\section{A possible Schwinger boson representation for the 
$su(M+1)$ algebra and its related $su(N,1)$ algebra}

In (I), we developed a possible boson representation for the 
$su(M+1)$ algebra, which is a natural extension of the 
Schwinger boson representation for the $su(2)$ algebra. 
First, we recapitulate its representation. 
Let us introduce a boson space which is constructed in terms 
of $(M+1)(N+1)$ kinds of boson operators : 
$({\hat a}_i, {\hat a}_i^*)$, $({\hat a}^p, {\hat a}^{p*})$, 
$({\hat b}_i^p, {\hat b}_i^{p*})$ and 
$({\hat b}, {\hat b}^*)$, where $i=1, 2, \cdots, M$ 
and $p=1, 2, \cdots, N$. 
In this space, the following bilinear forms are defined : 
\begin{subequations}\label{2-1}
\begin{eqnarray}
& &{\hat S}^i={\hat a}_i^*{\hat b}
+\sum_{p=1}^{N}{\hat a}^{p*}{\hat b}_i^p \ , \qquad
{\hat S}_i={\hat b}^*{\hat a}_i
+\sum_{p=1}^{N}{\hat b}_i^{p*}{\hat a}^p \ , \label{2-1a}\\
& &{\hat S}_i^j={\hat a}_i^*{\hat a}_j-\sum_{p=1}^N
{\hat b}_j^{p*}{\hat b}_i^p
+\delta_{ij}\left(\sum_{p=1}^N{\hat a}^{p*}{\hat a}^p
-{\hat b}^*{\hat b}\right) \ . \label{2-1b}
\end{eqnarray}
\end{subequations}
The set $({\hat S}^i, {\hat S}_i, {\hat S}_i^j)$ composes the 
$su(M+1)$ algebra : 
\begin{subequations}\label{2-2}
\begin{eqnarray}
& &{\hat S}_i^*={\hat S}^i \ , \qquad 
{\hat S}_j^{i*}={\hat S}_i^j \ , \label{2-2a}\\
& &[ {\hat S}^i , {\hat S}^j ]=0 \ , \qquad
[ {\hat S}^i , {\hat S}_j ]={\hat S}_i^j \ , \nonumber\\
& &[ {\hat S}_i^j , {\hat S}^k ]=\delta_{jk}{\hat S}^i
+\delta_{ij}{\hat S}^k \ , \nonumber\\
& &[ {\hat S}_i^j , {\hat S}_k^l ]=\delta_{jk}{\hat S}_i^l
-\delta_{il}{\hat S}_k^j \ .\label{2-2b}
\end{eqnarray}
\end{subequations}
In associating with the above set, we introduce an operator 
${\hat S}$ in the form 
\begin{equation}\label{2-3}
{\hat S}=\sum_{i=1}^M{\hat a}_i^*{\hat a}_i
-\sum_{p=1}^N{\hat a}^{p*}{\hat a}^p
-\sum_{i=1}^M\sum_{p=1}^N{\hat b}_i^{p*}{\hat b}_i^p
+{\hat b}^*{\hat b} \ .
\end{equation}
The operator ${\hat S}$ cannot be expressed in terms of any 
function for $({\hat S}^i, {\hat S}_i, {\hat S}_i^j)$ and 
satisfies 
\begin{equation}\label{2-4}
[ {\hat S}, {\hat S}^i]=[ {\hat S}, {\hat S}_i ]=
[ {\hat S} , {\hat S}_i^j ]=0 \ .
\end{equation}
The Casimir operator ${\hat \Gamma}_{su(M+1)}$, which 
commutes with $({\hat S}^i, {\hat S}_i, {\hat S}_i^j)$, 
is expressed as follows : 
\begin{eqnarray}\label{2-5}
{\hat \Gamma}_{su(M+1)}
&=&\sum_{i=1}^M({\hat S}^i{\hat S}_i+{\hat S}_i{\hat S}^i)
+\sum_{i, j=1}^M{\hat S}_j^i{\hat S}_i^j
-(M+1)^{-1}\left(\sum_{i=1}^M{\hat S}_i^i\right)^2 \nonumber\\
&=&2\left(\sum_{i=1}^M{\hat S}^i{\hat S}_i
+\sum_{j>i}{\hat S}_j^i{\hat S}_i^j\right) 
\nonumber\\
& &+\sum_{i=1}^M({\hat S}_i^i)^2
-(M+1)^{-1}\left(\sum_{i=1}^M{\hat S}_i^i\right)^2
+\sum_{i=1}^M(M-2i){\hat S}_i^i \ .
\end{eqnarray}

In the present boson space, we can define another set of the bilinear 
forms : 
\begin{subequations}\label{2-6}
\begin{eqnarray}
& &{\hat T}^p={\hat a}^{p*}{\hat b}^*
-\sum_{i=1}^M{\hat a}_i^*{\hat b}_i^{p*} \ , \qquad
{\hat T}_p={\hat b}{\hat a}^p
-\sum_{i=1}^M{\hat b}_i^p{\hat a}_i \ , \label{2-6a}\\
& &{\hat T}_q^p={\hat a}^{p*}{\hat a}^q
+\sum_{i=1}^M{\hat b}_i^{p*}{\hat b}_i^q
+\delta_{pq}\left(
\sum_{i=1}^M{\hat a}_i^*{\hat a}_i
+{\hat b}^*{\hat b}+(M+1)\right) \ .\label{2-6b}
\end{eqnarray}
\end{subequations}
The set $({\hat T}^p, {\hat T}_p,{\hat T}^p_q)$ composes the 
$su(N,1)$ algebra and satisfies the following relations : 
\begin{subequations}\label{2-7}
\begin{eqnarray}
& &{\hat T}_p^*={\hat T}^p \ , \qquad
{\hat T}_p^{q*}={\hat T}_q^p \ , \label{2-7a}\\
& &[ {\hat T}^p,{\hat T}^q ]=0 , \qquad
[ {\hat T}^p , {\hat T}_q ]=-{\hat T}_q^p \ , \nonumber\\
& &[ {\hat T}_q^p , {\hat T}^r ]=\delta_{qr}{\hat T}^p
+\delta_{pq}{\hat T}^r \ , \nonumber\\
& &[ {\hat T}_q^p , {\hat T}_s^r ]=\delta_{qr}{\hat T}_s^p
-\delta_{ps}{\hat T}_q^r \ .\label{2-7b}
\end{eqnarray}
\end{subequations}
We can easily verify the relation 
\begin{equation}\label{2-8}
[ {\hat S}, {\hat T}^p]=[ {\hat S}, {\hat T}_p ]=
[ {\hat S} , {\hat T}_q^p ]=0 \ .
\end{equation}
The Casimir operator ${\hat \Gamma}_{su(N,1)}$, which 
commutes with $({\hat T}^p, {\hat T}_p, {\hat T}_q^p)$, 
is expressed as follows : 
\begin{eqnarray}\label{2-9}
{\hat \Gamma}_{su(N,1)}
&=&-\sum_{p=1}^N({\hat T}^p{\hat T}_p+{\hat T}_p{\hat T}^p)
+\sum_{p, q=1}^N{\hat T}_q^p{\hat T}_p^q
-(N+1)^{-1}\left(\sum_{p=1}^N{\hat T}_p^p\right)^2 \nonumber\\
&=&-2\left(\sum_{p=1}^N{\hat T}^p{\hat T}_p
-\sum_{q>p}{\hat T}_p^q{\hat T}_q^p\right) 
\nonumber\\
& &+\sum_{p=1}^N({\hat T}_p^p)^2
-(N+1)^{-1}\left(\sum_{p=1}^N{\hat T}_p^p\right)^2
+\sum_{p=1}^N(N-2p){\hat T}_p^p \ .
\end{eqnarray}

The above is the recapitulation of (I) and, finally, 
a very important fact should be mentioned. 
The two sets $({\hat S}^i, {\hat S}_i,{\hat S}^j_i)$ and 
$({\hat T}^p, {\hat T}_p, {\hat T}_q^p)$ commute 
mutually for any components :
\begin{equation}\label{2-10}
[ {\hat S}^i , {\hat T}^p ]=0 \ , \ \ {\rm etc.}
\end{equation}

\section{A generalized Lipkin model}

We consider a nuclear shell model consisting of $(M+1)$ 
single-particle levels, each of which is specified by a 
quantum number $i\ (i=0, 1, 2, \cdots, M)$. 
The degeneracy of each level is $\Omega$-fold. 
Then, introducing a quantum number $m\ (m=-j, -j+1,\cdots, j-1, j\ :
j=\hbox{\rm half-integer}, \Omega=2j+1)$, 
the single-particle state can be specified by a set of the 
quantum numbers $(i, m)$. 
In other places, we use the notations $j$ and $m$ in a 
meaning different from the above, but, the confusion may not occur. 
For the fermion operators, we use particle creation and 
annihilation operator in the state $(i, m : i=1, 2,\cdots,M)$ 
by ${\maru \alpha}_{im}^*$ and ${\maru \alpha}_{im}$, respectively. 
For the state $(i=0, m)$, we use hole creation and annihilation 
operator ${\maru \beta}_{\widetilde m}^*$ and ${\maru \beta}_{\widetilde m}$, 
respectively, where the symbol ${\widetilde m}$ is used for abbreviating 
$(-1)^{j-m}{\maru T}_{-m}={\maru T}_{\widetilde m}$.

With the use of the above fermion operators, 
${\maru \alpha}_{im}^*$, ${\maru \alpha}_{im}$, 
${\maru \beta}_{\widetilde m}^*$ and ${\maru \beta}_{\widetilde m}$, 
we define the following bilinear form :
\begin{subequations}\label{3-1}
\begin{eqnarray}
& &
{\maru S}^i=\sum_m {\maru \alpha}_{im}^*{\maru \beta}_{\widetilde m}^* \ , 
\qquad
{\maru S}_i=\sum_m {\maru \beta}_{\widetilde m}{\maru \alpha}_{im} \ , 
\label{3-1a}\\
& &
{\maru S}_i^j=\sum_m {\maru \alpha}_{im}^*{\maru \alpha}_{jm}
-\delta_{ij}\left(\Omega-\sum_m{\maru \beta}_{\widetilde m}^*
{\maru \beta}_{\widetilde m}\right) \ .
\label{3-1b}
\end{eqnarray}
\end{subequations}
The symbol $j$ in the form (\ref{3-1b}) is used for specifying 
the $j$-th single-particle level. 
We know that the above set $({\maru S}^i, {\maru S}_i, {\maru S}_i^j)$ 
composes the $su(M+1)$ algebra and the properties are 
the same as those given in the relations (\ref{2-2a}) and 
(\ref{2-2b}). 
The Casimir operator ${\maru \Gamma}_{su(M+1)}$ is also 
of the same form as that given in the relation (\ref{2-5}). 
In addition to the above mathematical framework, conventionally, the simplest 
case, where there are $\Omega$ fermions, exactly the degeneracy of each level, 
has been investigated. 
Certainly, the case $M=1$ corresponds to the Lipkin model. 
In this sense, the above may be called as a generalized Lipkin model. 

However, we should note that there exists another way for the 
generalization of the Lipkin model. 
As was already mentioned, the total fermion number 
in the Lipkin model is equal to the degeneracy of each level. 
Keeping this relation, the above-mentioned model is generalized 
from the $su(2)$ algebra to the $su(M+1)$ algebra. 
Therefore, another way for the generalization is found in the 
case where the total fermion number is different of the 
degeneracy of each level. 
Then, in associating with the set 
$({\maru S}^i,{\maru S}_i, {\maru S}_i^j)$, 
we introduce an operator ${\maru N}$ in the form 
\begin{equation}\label{3-2}
{\maru N}=\left(\Omega-\sum_{m}
{\maru \beta}_{\widetilde m}^*{\maru \beta}_{\widetilde m}\right)
+\sum_{i=1}^M\left(
\sum_m{\maru \alpha}_{im}^*{\maru \alpha}_{im}\right) \ .
\end{equation}
The operator ${\maru N}$ represents the total fermion number 
and satisfies 
\begin{equation}\label{3-3}
[ {\maru N} , {\maru S}^i ]=[ {\maru N} , {\maru S}_i ]=
[ {\maru N} , {\maru S}_i^j ]=0 \ .
\end{equation}
It should be also 
noted that ${\maru N}$ cannot be expressed in terms of 
any function for 
$({\maru S}^i, {\maru S}_i, {\maru S}_i^j)$. 
In this paper, we will investigate the case where the 
total fermion number is different of the degeneracy of 
each level, even if the case of the $su(2)$ algebra.

Let us presuppose that, in the present fermion space, there exists the 
state $\kket{m}$ uniquely, which obeys the following conditions : 
\begin{equation}\label{3-4}
{\maru S}_i\kket{m}=0 \ , \qquad
{\maru S}_i^j\kket{m}=0\ . \quad
(j>i)
\end{equation}
Then, operating ${\maru S}_i^i$ on the both sides of the 
conditions (\ref{3-4}) and using the commutation relations, we have 
\begin{equation}\label{3-5}
{\maru S}_i\cdot{\maru S}_i^i\kket{m}=0 \ , \qquad
{\maru S}_i^j\cdot{\maru S}_i^i\kket{m}=0 \ .\quad
(j>i)
\end{equation}
Under the relations (\ref{3-5}), the presupposition of 
the existence of the unique $\kket{m}$ leads us to 
\begin{equation}\label{3-6}
{\maru S}_i^i\kket{m}=-\sigma_i\kket{m} \ . \quad
(i=1,2,\cdots, M)
\end{equation}
Here, $\sigma_i$ denotes $c$-number. 
The relation (\ref{3-6}) is nothing but the eigenvalue 
equation for the operator ${\maru S}_i^i$ and $\sigma_i$ 
is real. 
Further, operating ${\maru N}$ on the both sides 
of the conditions (\ref{3-4}), together with the relation 
(\ref{3-6}), we have 
\begin{subequations}
\begin{eqnarray}\label{3-7}
& &\sum_{m}{\maru \beta}_{\widetilde m}^*{\maru \beta}_{\widetilde m}
\kket{m}=n\kket{m} \ , \qquad\qquad\qquad\qquad\qquad\qquad
\qquad\qquad\qquad
\label{3-7a}\\
& &\sum_{m}{\maru \alpha}_{im}^*{\maru \alpha}_{im}
\kket{m}=n_i\kket{m} \ , \label{3-7b}
\end{eqnarray}
\end{subequations}
\begin{eqnarray}
& &N=\Omega-n+\sum_{i=1}^Mn_i \ , \qquad
\Omega-n=(M+1)^{-1}\left(N+\sum_{j=1}^M\sigma_j \right) \ , 
\label{3-8}\\
& &\sigma_i=\Omega-n-n_i \ , \qquad
n_i=(M+1)^{-1}\left(N+\sum_{j=1}^M\sigma_j\right)-\sigma_i \ .
\label{3-9}
\end{eqnarray}
The relation (\ref{3-8}) tells us that we are interested in  
a system with $N\ (=\Omega-n+\sum_{i=1}^M n_i)$ fermion 
number. 
The above consideration means that the state $\kket{m}$ is 
specified by 
\begin{equation}\label{3-10}
\kket{m}=\kket{(\gamma); N, \sigma_1, \cdots,\sigma_M}\ \ \ \ 
{\rm or}\ \ \ \ 
\kket{(\gamma); n, n_1, \cdots,n_M} \ .
\end{equation}
Here, $(\gamma)$ denotes a set of the quantum numbers which 
do not relate directly to the present algebra.

Now, let us investigate properties of the quantum numbers 
specifying the state (\ref{3-10}). 
First, we note the relations 
\begin{equation}\label{3-11}
\bbra{m}{\maru S}_i\cdot{\maru S}^i\kket{m} \ge 0 \ , \qquad
\bbra{m}{\maru S}_i^j\cdot{\maru S}_j^i\kket{m} \ge 0 \ . 
\ \ (j>i)
\end{equation}
The conditions (\ref{3-11}) give us the restrictions 
\begin{eqnarray}
& &\sigma_i \ge 0 \ , \quad {\rm i.e.,}\quad
n+n_i \le \Omega \ , \label{3-12}\\
& &\sigma_i\le \sigma_j \ , \quad{\rm i.e.,}\quad
n_i \ge n_j \ . \ \ (j>i) \label{3-13}
\end{eqnarray}
The restrictions (\ref{3-12}) and (\ref{3-13}) are summarized as 
\begin{equation}\label{3-14}
\Omega-n \ge n_1 \ge n_2 \ge \cdots \ge n_M \ .
\end{equation}

It may be self-evident that the state $\kket{m}$ is the 
eigenstate of the Casimir operator ${\maru \Gamma}_{su(M+1)}$ 
with the eigenvalue $\gamma_{su(M+1)}^{(f)}$, which is given as 
\begin{equation}\label{3-15}
\gamma_{su(M+1)}^{(f)}
=\sum_{i=1}^M\sigma_i^2
-(M+1)^{-1}\left(\sum_{i=1}^M\sigma_i\right)^2
-\sum_{i=1}^M(M-2i)\sigma_i \ .
\end{equation}
The above form is obtained through the relation (\ref{2-5}) 
replaced ${\hat S}_i^i$ with ${\maru S}_i^i$ and Eq.(\ref{3-6}). 
If $c$, $e$, $L$, $X_c$, $X^e$ and $X^L$ in the relations 
(\ref{A-16}) and (\ref{A-17}) read $i$, $k$, $M$, $-\sigma_i$, 
$-\sigma^k$ and $-\sigma^M$, respectively, the form (\ref{3-15}) 
can be also written as follows : 
\begin{eqnarray}
& &\gamma_{su(M+1)}^{(f)}
=\sum_{k=1}^M [k/(k+1)]\sigma^k[\sigma^k+(k+1)] \ , \label{3-16}\\
& &\sigma^k=
\cases{ -(1/k){\displaystyle \sum_{i=1}^k}
\sigma_i+\sigma_{k+1}\ , \quad
(k=1,2,\cdots, M-1) \cr
(1/M){\displaystyle \sum_{i=1}^M}\sigma_i \ . \qquad\qquad (k=M)}
\label{3-17}
\end{eqnarray}
Then, the state $\kket{m}$ can be also specified as 
\begin{equation}\label{3-18}
\kket{m}=\kket{(\gamma); N,\sigma^1, \cdots ,\sigma^M} \ .
\end{equation}
Under the restriction (\ref{3-13}), we can show that 
$\gamma_{su(M+1)}^{(f)}$ is positive-definite.

Thus, we learned the properties of the state $\kket{m}$ 
which obeys the conditions (\ref{3-4}). 
The state $\kket{m}$ is specified by $(M+1)$ quantum numbers. 
By regarding $\kket{m}$ as the intrinsic state and by operating 
${\maru S}^i$ and ${\maru S}_j^i\ (j>i)$ in 
an appropriate order on $\kket{m}$, we are able to obtain 
the orthogonal set for the $su(M+1)$ algebra. 
The number of ${\maru S}^i$ and ${\maru S}_j^i\ (j>i)$ 
are $M$ and $(M^2-M)/2$, respectively and totally, 
$(M^2+M)/2$. 
Then, the orthogonal set is specified by $(M+1)(M+2)/2 
\ (=(M^2+M)/2+(M+1))$ quantum numbers.

\section{Construction of the intrinsic state in the boson space}

Now, let us investigate the intrinsic state playing 
the same role as that of $\kket{m}$ shown in \S 3 in the 
boson space defined in \S 2. 
First, we impose the following conditions to the state $\ket{m}$ :
\begin{subequations}\label{4-1}
\begin{eqnarray}
& &{\hat S}_i\ket{m}=0 \ , \qquad
{\hat S}_i^j\ket{m}=0 \ , \quad (j>i) \label{4-1a}\\
& &{\hat T}_p\ket{m}=0 \ , \qquad
{\hat T}_q^p\ket{m}=0 \ . \quad (q>p) \label{4-1b}
\end{eqnarray}
\end{subequations}
For a moment, the conditions (\ref{4-1a}) and (\ref{4-1b}) 
should be regarded as the supposition, which is in the same 
situation as that in \S 3. 
Further, let the state $\ket{m}$ contain $(M+1)$ quantum numbers. 
Then, by operating ${\hat S}^i$, 
${\hat S}_j^i \ (j>i)$, ${\hat T}^p$ and ${\hat T}_p^q\ (q>p)$ 
appropriately on the state $\ket{m}$, we are able to obtain the 
orthogonal set for the present algebras. 
If it is possible, the total number of the quantum numbers 
is given by $(N^2+N)/2+(M^2+M)/2+(M+1)$. 
On the other hand, our present boson space consists of 
$(M+1)(N+1)$ kinds of bosons and, then, we have the relation 
\begin{equation}\label{4-2}
(N^2+N)/2+(M^2+M)/2+(M+1)=(M+1)(N+1) \ .
\end{equation}
The above relation gives us 
\begin{equation}\label{4-3}
N=M\ , \qquad N=M+1 \ .
\end{equation}
In this paper, we will treat the case $N=M$.

In \S 3, starting from the presupposition of the existence 
of the unique $\kket{m}$, we showed its various properties, 
but, we did not prove the existence. 
In this section, in the same idea as that in \S 3, we will start 
with the presupposition of the unique $\ket{m}$, but, the proof 
of the existence and its explicit form are given. 
From the condition ${\hat S}_i\ket{m}={\hat T}_p\ket{m}=0$ 
for $i, p=1, 2, \cdots, M$ shown in Eqs.(\ref{4-1a}) 
and (\ref{4-1b}), we can conclude that the intrinsic state 
$\ket{m}$ contains only ${\hat b}_i^{p*}\ (i, p=1, 2, \cdots, M)$ 
and ${\hat b}^*$. 
Therefore, it may be enough to consider the conditions 
${\hat S}_i^j\ket{m}={\hat T}_q^p\ket{m}=0\ (j>i, q>p)$. 
In order to obtain the state $\ket{m}$ which satisfies the 
above conditions, we introduce the following operator :
\begin{equation}\label{4-4}
{\hat B}_r^*=
\left|
\matrix{ {\hat b}_M^{1*} & {\hat b}_M^{2*} & \cdots & {\hat b}_M^{r*} \cr
         {\hat b}_{M-1}^{1*} & {\hat b}_{M-1}^{2*} &
             \cdots & {\hat b}_{M-1}^{r*} \cr
          \cdots & \cdots & \cdots & \cdots \cr
         {\hat b}_{M-r+1}^{1*} & {\hat b}_{M-r+1}^{2*} &
          \cdots & {\hat b}_{M-r+1}^{r*} 
         }
         \right|
\ . \quad (r=1, 2, \cdots, M)
\end{equation}
For example, ${\hat B}_1^*$ and ${\hat B}_2^*$ are given as 
\begin{equation}\label{4-5}
{\hat B}_1^*={\hat b}_M^{1*} \ , \qquad
{\hat B}_2^*=\left|\matrix{ {\hat b}_M^{1*} & {\hat b}_M^{2*} \cr
                            {\hat b}_{M-1}^{1*} & {\hat b}_{M-1}^{2*}
                           }\right|
={\hat b}_M^{1*}{\hat b}_{M-1}^{2*}-{\hat b}_{M-1}^{1*}{\hat b}_M^{2*} \ .
\end{equation}
The operator ${\hat B}_r^*$ obeys the relations 
\begin{subequations}\label{4-6}
\begin{eqnarray}
& &[ {\hat S}_i^j , {\hat B}_r^* ] =0 \ , \quad (j>i) \label{4-6a}\\
& &[ {\hat T}_q^p , {\hat B}_r^* ] =0 \ . \quad (q>p) \label{4-6b}
\end{eqnarray}
\end{subequations}
The above relations are proved in Appendix B. 
As for a possible form for $\ket{m}$, we adopt the 
following form : 
\begin{equation}\label{4-7}
\ket{m}=({\hat B}_1^*)^{\nu_M}({\hat B}_2^*)^{\nu_{M-1}}\cdots
({\hat B}_r^*)^{\nu_{M-r+1}}\cdots ({\hat B}_M^*)^{\nu_1}
({\hat b}^*)^\nu \ket{0} \ .
\end{equation}
Since ${\hat S}_i^j\ket{0}={\hat T}_q^p\ket{0}=0$ and there exist the 
properties (\ref{4-6a}) and (\ref{4-6b}), the following 
relations may be self-evident : 
\begin{equation}\label{4-8}
{\hat S}_i^j \ket{m}=0 \ , \quad
(j>i) \ , \qquad
{\hat T}_q^p \ket{m}=0 \ . \quad (q>p)
\end{equation}
Certainly, we can learn that there exists a state, which satisfies 
the relations (\ref{4-1a}) and (\ref{4-1b}), in the present 
boson space.

Our next task is to show that the state $\ket{m}$ given 
in Eq.(\ref{4-7}) is the eigenstate of the operators 
${\hat S}_i^i$ and ${\hat T}_p^p$. 
For this aim, the following relations, which are proved in 
Appendix B, are useful : 
\begin{subequations}\label{4-9}
\begin{eqnarray}
& & [ {\hat S}_i^i , {\hat B}_r^* ] 
=\cases{ 0 \ , \qquad (r=1, 2, \cdots, M-i-1) \cr
         -{\hat B}_r^* \ ,\quad (r=M-i, \cdots, M)
         }\label{4-9a}\\
& & [ {\hat T}_p^p , {\hat B}_r^* ] 
=\cases{ 0 \ , \qquad (r=1, 2, \cdots, p-1) \cr
         {\hat B}_r^* \ ,\quad (r=p, p+1, \cdots, M)
         }\label{4-9b}
\end{eqnarray}
\end{subequations}
Then, it easily shown that the state $\ket{m}$ is the 
eigenstate of ${\hat S}_i^i$ and ${\hat T}_p^p$ : 
\begin{subequations}\label{4-10}
\begin{eqnarray}
& & {\hat S}_i^i\ket{m}=-s_i\ket{m} \ , \qquad
s_i=\sum_{k=1}^i \nu_k+\nu \ , \label{4-10a}\\
& & {\hat T}_p^p\ket{m}=+t_p\ket{m} \ , \qquad
t_p=(M+1)+\sum_{k=1}^{M-p+1} \nu_k+\nu \nonumber\\
& &\qquad\qquad\qquad\qquad\qquad\quad\ 
=(M+1)+s_{M-p+1}  \ . \label{4-10b}
\end{eqnarray}
\end{subequations}
Since $s_i-s_{i-1}=\sum_{k=1}^i \nu_k-\sum_{k=1}^{i-1} \nu_k
=\nu_i \ (i \ge 2)$ and $s_1=\nu_1+\nu$, 
we have 
\begin{eqnarray}\label{4-11}
\ket{m}=& &({\hat B}_1^*)^{s_M-s_{M-1}}({\hat B}_2^*)^{s_{M-1}-s_{M-2}}
\cdots({\hat B}_{M-k+1}^*)^{s_k-s_{k-1}} \nonumber\\
& &\cdots
({\hat B}_{M-1}^*)^{s_2-s_{1}}({\hat B}_M^*)^{s_1-\nu}
({\hat b}^*)^{\nu}\ket{0} \ .
\end{eqnarray}
Further, the state (\ref{4-11}) can be rewritten in another form : 
\begin{subequations}\label{4-12}
\begin{eqnarray}
& &\ket{m}=({\hat A}_M^*)^{s_1-\nu}\ket{s_1, s_2, \cdots, s_M} \ , 
\label{4-12a}\\
& &\ket{s_1, s_2,\cdots, s_M}
=({\hat B}_1^*)^{s_M-s_{M-1}}({\hat B}_2^*)^{s_{M-1}-s_{M-2}}
\nonumber\\
& &\qquad\qquad\qquad\qquad
\cdots({\hat B}_{M-k+1}^*)^{s_k-s_{k-1}}
\cdots
({\hat B}_{M-1}^*)^{s_2-s_{1}}({\hat b}^*)^{s_1}\ket{0} \ .
\label{4-12b}
\end{eqnarray}
\end{subequations}
Here, the operator ${\hat A}_M^*$ is defined in the following 
form : 
\begin{equation}\label{4-13}
{\hat A}_M^*={\hat B}_M^*{\hat b}
+\sum_{l=1}^M {\hat B}_M^*(l){\hat a}_l \ .
\end{equation}
Since ${\hat a}_l\ket{s_1, s_2, \cdots, s_M}=0$, 
the state (\ref{4-12}) returns to the state (\ref{4-11}) for any 
form of ${\hat B}_M^*(l)$. 
However, we choose ${\hat B}_M^*(l)$ in the form that 
${\hat A}_M^*$ commutes with all the generators : 
\begin{equation}\label{4-14}
[ {\hat S}^i , {\hat A}_M^* ] =
[ {\hat S}_i , {\hat A}_M^* ] =
[ {\hat S}_i^j , {\hat A}_M^* ] =
[ {\hat T}^p , {\hat A}_M^* ] =
[ {\hat T}_p , {\hat A}_M^* ] =
[ {\hat T}_p^q , {\hat A}_M^* ] =0 \ .
\end{equation}
The operator ${\hat B}_M^*(l)$ which satisfies the condition 
(\ref{4-14}) is given in the form that all the elements 
of the $(M-l+1)$-th row in ${\hat B}_r^*$ given in Eq.(\ref{4-4}) 
for $r=M$, ${\hat b}_l^{r*}\ (r=1, 2, \cdots, M)$, 
are replaced with ${\hat a}^r$ : 
\begin{equation}\label{4-15}
{\hat B}_M^*(l)=
\left|
\matrix{ {\hat b}_M^{1*} & {\hat b}_M^{2*} & \cdots & {\hat b}_M^{M*} \cr
         \cdots & \cdots & \cdots & \cdots \cr
         {\hat b}_{M-l+2}^{1*} & {\hat b}_{M-l+2}^{2*} &
          \cdots & {\hat b}_{M-l+2}^{M*} \cr
         {\hat a}^{1} & {\hat a}^{2} & \cdots & {\hat a}^{M} \cr
         {\hat b}_{M-l}^{1*} & {\hat b}_{M-l}^{2*} &
          \cdots & {\hat b}_{M-l}^{M*} \cr
        \cdots & \cdots & \cdots & \cdots \cr
         {\hat b}_{1}^{1*} & {\hat b}_{1}^{2*} &
          \cdots & {\hat b}_{1}^{M*} 
         }
         \right|
\ . \quad (l=1, 2, \cdots, M)
\end{equation}
The proof is sketched in Appendix B. 
The property (\ref{4-14}) will play an essential role in \S 7. 
All $\nu_i$ in the state (\ref{4-7}) should be positive and, 
then, the state (\ref{4-11}) and its reform (\ref{4-12}) give 
us the following relations :
\begin{eqnarray}
& &0 \le s_1 \le s_2 \le \cdots \le s_M \ , \label{4-16}\\
& &0 \le \nu \le s_1 \ . \label{4-17}
\end{eqnarray}
Thus, we could prove the existence of the state $\ket{m}$ 
which obeys the conditions (\ref{4-1a}) and (\ref{4-1b}) 
with the explicit form. 
In the same meaning as that mentioned in \S 3, $\ket{m}$ can be 
regarded as the intrinsic state. 
The state $\ket{m}$ is the eigenstate of ${\hat S}$ introduced 
in Eq.(\ref{2-3}) and its eigenvalue $S$ is given as 
\begin{subequations}\label{4-18}
\begin{equation}\label{4-18a}
S=\sum_{i=1}^M (\nu-s_i)+\nu \ .
\end{equation}
Therefore, we have 
\begin{equation}\label{4-18b}
\nu=(M+1)^{-1}\left(S+\sum_{i=1}^M s_i\right) \ , \quad
{\rm i.e.,}\ \ 
s_1-\nu=s_1-(M+1)^{-1}\left(S+\sum_{i=1}^M s_i \right) \ .
\end{equation}
\end{subequations}
Then, $\ket{m}$ is specified as 
\begin{equation}\label{4-19}
\ket{m}=\ket{S, s_1, s_2, \cdots, s_M} \ .
\end{equation}
The relation (\ref{4-17}) gives us the relation 
\begin{equation}\label{4-20}
-\sum_{i=1}^M s_i \le S \le (M+1)s_1-\sum_{i=1}^M s_i \ .
\end{equation}

Finally, we will discuss the eigenvalues of the Casimir 
operators ${\hat \Gamma}_{su(M+1)}$ and 
${\hat \Gamma}_{su(M,1)}$. 
The eigenvalues $\gamma_{su(M+1)}^{(b)}$ and 
$\gamma_{su(M,1)}^{(b)}$ for the state $\ket{m}$ are given as 
\begin{subequations}\label{4-21}
\begin{eqnarray}
& &\gamma_{su(M+1)}^{(b)}
=\sum_{i=1}^M s_i^2-(M+1)^{-1}\left(\sum_{i=1}^M s_i\right)^2
-\sum_{i=1}^M(M-2i)s_i \ , \label{4-21a}\\
& &\gamma_{su(M,1)}^{(b)}
=\sum_{p=1}^M t_p^2-(M+1)^{-1}\left(\sum_{p=1}^M t_p\right)^2
-\sum_{p=1}^M(M-2p)t_p \ . \label{4-21b}
\end{eqnarray}
\end{subequations}
Of course, in the present, the case $N=M$ is treated. 
In the same way as that mentioned in \S 3, both are expressed as 
\begin{subequations}\label{4-22}
\begin{eqnarray}
& &\gamma_{su(M+1)}^{(b)}
=\sum_{k=1}^{M-1} [k/(k+1)] s^k [s^k+(k+1)]
+[M/(M+1)]s^M[s^M+(M+1)] \ , \nonumber\\
& & \label{4-22a}\\
& &\gamma_{su(M,1)}^{(b)}
=\sum_{r=1}^{M-1} [r/(r+1)] t^r [t^r+(r+1)]
+[M/(M+1)]t^M[t^M-(M+1)] \ , \nonumber\\
& &\label{4-22b}
\end{eqnarray}
\vspace{-0.6cm}
\end{subequations}
\begin{subequations}\label{4-23}
\begin{eqnarray}
& & s^k=\cases{\displaystyle 
-(1/k)\sum_{i=1}^k s_i +s_{k+1} \ , \quad
(k=1, 2, \cdots, M-1) \cr
\displaystyle               (1/M)\sum_{i=1}^M s_i \ , \qquad
(k=M)} \qquad\qquad\qquad\qquad
\label{4-23a}\\
& & t^r=\cases{\displaystyle 
(1/r)\sum_{p=1}^r t_p -t_{r+1} \ , \quad
(r=1, 2, \cdots, M-1) \cr
\displaystyle               (1/M)\sum_{p=1}^M t_p \ , \qquad
(r=M)} \label{4-23b}
\end{eqnarray}
\end{subequations}
The negative sign in the form (\ref{4-23b}) characterizes 
the $su(M,1)$ algebra. 
As is clear from the relation (\ref{4-10b}), 
$t^r$ can be expressed in terms of $s^i$. 
The state $\ket{m}$ is also specified as 
\begin{equation}\label{4-24}
\ket{m}=\ket{S, s^1, s^2,\cdots, s^M} \ .
\end{equation}

\section{The generalized Lipkin model in the Schwinger boson 
representation}

In \S 3, we formulate the generalized Lipkin model 
in the fermion space and showed various properties of the 
intrinsic state $\kket{m}\ (=\kket{(\gamma); n, n_1, \cdots, n_M})$. 
Further, in \S 4, the intrinsic state for $su(M+1)$ algebraic model 
(the $su(M,1)$ algebraic model), $\ket{m}\ (=\ket{S, s_1, \cdots, s_M})$ 
was investigated. 
The aim of this section is to reinvestigate the intrinsic 
state $\ket{m}$ in terms of $(n, n_1, \cdots, n_M)$ 
or $(N, n_1, \cdots, n_M)$.

First, we note two operators ${\hat S}$ and ${\maru N}$. 
Both are commuted with any generator of the $su(M+1)$ algebras 
expressed in terms of boson and fermion operators and 
${\hat S}$ is commuted with any generator of the $su(M,1)$ 
algebra. 
However, there exists an essential difference between 
${\hat S}$ and ${\maru N}$ : 
${\hat S}$ is not positive-definite, but, ${\maru N}$ is 
positive-definite. 
Therefore, it may be impossible to regard ${\hat S}$ 
as the counterpart of ${\maru N}$ and, then, we treat total 
fermion number $N$ as a parameter in the Schwinger boson 
representation. 
However, from the relations (\ref{3-6}) and (\ref{4-10a}), 
it may be permitted to set up 
\begin{equation}\label{5-1}
\sigma_i=s_i \ . \quad (i=1, 2, \cdots, M)
\end{equation}
Then, we introduce the quantities $(\Omega-n)$ and $n_i$ into 
the Schwinger boson representation through the relations 
\begin{subequations}\label{5-2}
\begin{eqnarray}
\Omega-n &=& (M+1)^{-1}\left(N+\sum_{j=1}^M s_j\right) \ , 
\label{5-2a}\\
n_i&=&(M+1)^{-1}\left(N+\sum_{j=1}^M s_j\right)-s_i \ .
\label{5-2b}
\end{eqnarray}
\end{subequations}
Inversely, we have 
\begin{subequations}\label{5-3}
\begin{eqnarray}
& &N=\Omega-n+\sum_{i=1}^M n_i \ , \label{5-3a}\\
& &s_i=\Omega-n-n_i \ . \label{5-3b}
\end{eqnarray}
\end{subequations}
With the use of the relations (\ref{5-2}), the inequality (\ref{4-16}) 
can be rewritten as 
\begin{equation}\label{5-4}
\Omega-n \ge n_1 \ge n_2 \ge \cdots \ge n_M\ .
\end{equation}
The above is nothing but the inequality (\ref{3-14}). 
Further, the inequality (\ref{4-20}) is rewritten 
in the form 
\begin{equation}\label{5-5}
N-(M+1)(\Omega-n) \le S \le N-(M+1)n_1 \ .
\end{equation}

The relations (\ref{4-18a}) and (\ref{5-2}) give us 
\begin{eqnarray}
& &s_1-\nu=(N-S)/(M+1)-n_1 \ , \label{5-6}\\
& &s_k-s_{k-1}=n_k-n_{k-1} \ .\quad
(k=2, 3, \cdots, M) \label{5-7}
\end{eqnarray}
Then, the state $\ket{m}$ shown in the relation (\ref{4-12}) can be 
re-expressed in the form 
\begin{subequations}\label{5-8}
\begin{eqnarray}
& & \ket{m}=({\hat A}_M^*)^{(N-S)/(M+1)-n_1}
\dket{s_1, s_2, \cdots, s_M} \ , 
\label{5-8a}\\
& & \dket{s_1, s_2, \cdots, s_M}
=({\hat B}_1^*)^{n_{M-1}-n_M}({\hat B}_2^*)^{n_{M-1}-n_{M-1}}\nonumber\\
& &\qquad\qquad\qquad\qquad
\cdots({\hat B}_{M-k+1}^*)^{n_{k-1}-n_k}\cdots
({\hat B}_{M-1}^*)^{n_{1}-n_2}({\hat b}^*)^{\Omega-n-n_{1}}
\ket{0} \ . \qquad\label{5-8b}
\end{eqnarray}
\end{subequations}
The above shows that, with the use of the quantum numbers specifying 
the intrinsic state of the generalized Lipkin model for the 
fermion system, the explicit form is given in the Schwinger 
boson representation. 
It should be noted that, for a given set $(n, n_1, n_2, \cdots, n_M)$, 
there exist many states which are orthogonal with one another 
for different values of $S$ obeying the inequality (\ref{5-5}). 
The quantities $s^k$ and $t^r$ given in the relations 
(\ref{4-23a}) and (\ref{4-23b}), respectively, are 
expressed as 
\begin{subequations}\label{5-9}
\begin{eqnarray}
& &s^k=\cases{
\displaystyle (1/k)\sum_{i=1}^k n_i-n_{k+1} \ , \quad (k=1, 2, \cdots, M-1) \cr
\displaystyle \Omega-n-(1/M)\sum_{i=1}^M n_i \ , \quad (k=M)
} \label{5-9a}\\
& &t^r=\cases{
\displaystyle n_{M-r}-(1/r)\sum_{i=M-r+1}^M n_i \ , 
\quad (r=1, 2, \cdots, M-1) \cr
\displaystyle M+1+\Omega-n-(1/M)\sum_{i=1}^M n_i \ , \quad (r=M)
} \label{5-9b}
\end{eqnarray}
\end{subequations}
The inequality (\ref{5-4}) tells us that all $s^k$ and $t^r$ are 
positive.

\section{General scheme for constructing orthogonal set 
for the irreducible representation}

Until the previous section, we have investigated the method 
how to construct the intrinsic state $\ket{m}$. 
Then, our next task is to find the orthogonal set built on 
the state $\ket{m}$ by operating ${\hat S}^i$, ${\hat S}_j^i$, 
${\hat T}^p$ and ${\hat T}_p^q$ as the raising operators. 
Here, $i=1, 2, \cdots, M$, $p=1, 2, \cdots, M$, $j>i$ and 
$q>p$. 
The problem is to determine the ordering the operation of 
${\hat S}^i$, ${\hat S}_j^i$, 
${\hat T}^p$ and ${\hat T}_p^q$. 
Of course, there does not exist any problem between 
$({\hat S}^i , {\hat S}_j^i)$ and 
$({\hat T}^p , {\hat T}_p^q)$, because these commute with each other. 
In the sense of the full use of the raising operators, the present 
idea is similar to a method developed by Moshinsky for the shell 
model.\cite{M}

First, we investigate the operation of ${\hat S}^i \ 
(i=1, 2, \cdots, M)$. 
For this aim, let us define a set of the operators 
${\hat {\mib S}}^i$ expressed in terms of the linear 
combinations for ${\hat S}^l$ : 
\begin{equation}\label{6-1}
{\hat{\mib S}}^i=\sum_{l=1}^M{\hat S}^l{\hat U}_{il} \ .\quad
(i=1, 2, \cdots, M)
\end{equation}
Here, ${\hat U}_{il}$ is a function of ${\hat S}_j^i\ 
(j>i=1, 2, \cdots, M-1)$ and $({\hat S}_i^i-{\hat S}_1^1)\ 
(i=2, 3, \cdots, M)$. 
Under an appropriate choice of the coefficients ${\hat U}_{il}$, 
let ${\hat{\mib S}}^i$ satisfy the following relations : 
\begin{subequations}\label{6-2}
\begin{eqnarray}
& & [ {\hat S}_i^j , {\hat{\mib S}}^k ] 
=\sum_{m>l=1,\cdots,M-1} {\hat V}_{ijk,lm}{\hat S}_l^m \ , \quad 
(j>i=1, 2, \cdots, M-1) \label{6-2a}\\
& & [ {\hat S}_i^i , {\hat{\mib S}}^k ] 
=(1+\delta_{ik}){\hat{\mib S}}^k \ , \quad
(i=1, 2, \cdots, M) \label{6-2b}\\
& & [ {\hat{\mib S}}^i , {\hat{\mib S}}^j ] =0 \ . \label{6-2c}
\end{eqnarray}
\end{subequations}
Here, generally, ${\hat V}_{ijk,lm}$ is operators. 
For example, ${\hat{\mib S}}^1$, ${\hat{\mib S}}^2$ and 
${\hat{\mib S}}^3$ are given in the form 
\begin{eqnarray}
{\hat{\mib S}}^1&=&
{\hat S}^1 \ , \label{6-3}\\
{\hat{\mib S}}^2&=&{\hat S}^1\cdot{\hat S}_2^1+{\hat S}^2\cdot
({\hat S}_2^2-{\hat S}_1^1) \ , \label{6-4}\\
{\hat{\mib S}}^3&=&
{\hat S}^1\cdot\left[{\hat S}_2^1{\hat S}_3^2+
{\hat S}_3^1(({\hat S}_3^3-{\hat S}_1^1)-({\hat S}_2^2-{\hat S}_1^1))
\right] 
+{\hat S}^2\cdot{\hat S}_3^2\left[
({\hat S}_3^3-{\hat S}_1^1)-1\right]\nonumber\\
& &+{\hat S}^3\cdot\left[({\hat S}_3^3-{\hat S}_1^1)
-({\hat S}_2^2-{\hat S}_1^1)\right]\left[({\hat S}_3^3
-{\hat S}_1^1)-1\right] \ .\label{6-5}
\end{eqnarray}
With the use of the operator ${\hat{\mib S}}^i$, we define the 
following state : 
\begin{subequations}\label{6-6}
\begin{eqnarray}
& &\ket{m(1)}={\hat{\mib C}}_1(m_1^{(1)}, \cdots, m_M^{(1)})\ket{m} \ , 
\label{6-6a}\\
& &{\hat{\mib C}}_1(m_1^{(1)}, \cdots, m_M^{(1)})
=({\hat{\mib S}}^1)^{m_1^{(1)}}\cdots
({\hat{\mib S}}^M)^{m_M^{(1)}} \ . \label{6-6b}
\end{eqnarray}
\end{subequations}
The relation (\ref{6-2b}) shows us that the state $\ket{m(1)}$ 
is an eigenstate of ${\hat S}_i^i$ : 
\begin{equation}\label{6-7}
{\hat S}_i^i\ket{m(1)}
=-\left(s_i-m_i^{(1)}-\sum_{k=1}^Mm_k^{(1)}\right) \ket{m(1)} \ .
\quad (i=1, 2, \cdots , M)
\end{equation}
Therefore, the set $(\ket{m(1)})$ forms an orthogonal set and 
the relation (\ref{6-2c}) tells us that the ordering of 
${\hat{\mib S}}^i$ in ${\hat{\mib C}}_1$ is arbitrary. 
Further, we note the following relations : 
\begin{subequations}\label{6-8}
\begin{eqnarray}
& &{\hat S}_i^j\ket{m(1)}=0 \ , \qquad (j>i=1, 2, \cdots, M-1) 
\label{6-8a}\\
& &({\hat S}_i^i-{\hat S}_1^1)\ket{m(1)}
=-[(s_i-m_i^{(1)})-(s_1-m_1^{(1)})]\ket{m(1)} \ . \ \ 
(i=2, 3, \cdots, M) \qquad\quad
\label{6-8b}
\end{eqnarray}
\end{subequations}
The set $({\hat S}_j^i-\delta_{ij}{\hat S}_1^1)\ 
(i, j=1, 2, \cdots, M)$ forms the $su(M)$ algebra as 
a sub-algebra of the starting $su(M+1)$ algebra. 
The relations (\ref{6-8a}) and (\ref{6-8b}) show us that 
the state $\ket{m(1)}$ is the intrinsic state of the $su(M)$ 
algebra and ${\hat S}_j^i\ (j>i=1, 2, \cdots, M-1)$ is 
the raising operator on the intrinsic state $\ket{m(1)}$. 
From the above consideration, it may be concluded that the 
operation of ${\hat S}^i$ on the state $\ket{m}$ was 
finished in the form (\ref{6-6}).

Next task is how to operate ${\hat S}_j^i\ (j>i=1, 2, \cdots, M-1)$ 
on the state $\ket{m(1)}$. 
We already mentioned that the set $({\hat S}_j^i-\delta_{ij}
{\hat S}_1^1)\ (i, j=1, 2, \cdots, M)$ forms the $su(M)$ 
algebra and if its generators are decomposed to 
$({\hat S}_i^1, {\hat S}_1^i, (i=2,3,\cdots,M), 
{\hat S}_2^2-{\hat S}_1^1, {\hat S}_j^i-\delta_{ij}{\hat S}_2^2, 
(i,j=2,3,\cdots M))$, the set 
$({\hat S}_j^i-\delta_{ij}{\hat S}_2^2)\ (i, j=2, 3, \cdots, M)$ 
forms the $su(M-1)$ algebra. 
Under the above note, we consider the operation of 
${\hat S}_i^1\ (i=2, 3, \cdots, M)$ on the state $\ket{m(1)}$. 
In the same form as that shown in Eq.(\ref{6-1}), we introduce 
the operator ${\hat{\mib S}}_i^1$ in the form 
\begin{equation}\label{6-9}
{\hat{\mib S}}_i^1=\sum_{l=2}^M{\hat S}_l^1{\hat U}_{il}^{(1)} \ .
\quad (i=2, 3, \cdots, M)
\end{equation}
Here, ${\hat U}_{il}^{(1)}$ is a function of ${\hat S}_j^i\ 
(j>i=2, 3, \cdots, M-1)$ and $({\hat S}_i^i-{\hat S}_2^2)\ 
(i=3, 4, \cdots, M)$. 
In parallel to the relations (\ref{6-2}), ${\hat{\mib S}}_i^1$ 
is regarded as the operator satisfying the relations 
\begin{subequations}\label{6-10}
\begin{eqnarray}
& &[ {\hat S}_i^j , {\hat{\mib S}}_k^1 ]
=\sum_{m>l=2,\cdots,M-1}{\hat V}_{ijk,lm}^{(1)}{\hat S}_l^m \ , 
\quad (j>i=2, 3, \cdots, M-1) 
\label{6-10a}\\
& &[ {\hat S}_i^i-{\hat S}_1^1 , {\hat{\mib S}}_k^1 ]
=(1+\delta_{ik}){\hat{\mib S}}_k^1\ , \quad
(i=2, 3, \cdots, M)
\label{6-10b}\\
& &[ {\hat{\mib S}}_i^1 , {\hat{\mib S}}_j^1 ]=0 \ .
\label{6-10c}
\end{eqnarray}
\end{subequations}
For example, ${\hat {\mib S}}_2^1$ and ${\hat{\mib S}}_3^1$ are 
given as 
\begin{eqnarray}
& &{\hat{\mib S}}_2^1={\hat S}_2^1 \ , \label{6-11}\\
& &{\hat{\mib S}}_3^1={\hat S}_2^1\cdot{\hat S}_3^2
+{\hat S}_3^1\cdot({\hat S}_3^3-{\hat S}_2^2) \ .
\label{6-12}
\end{eqnarray}
With the use of the operator ${\hat{\mib S}}_i^1$, we define 
the following state :
\begin{subequations}\label{6-13}
\begin{eqnarray}
& &\ket{m(2)}={\hat{\mib C}}_2(m_2^{(2)}, \cdots, m_M^{(2)})
\ket{m(1)} \ , \label{6-13a}\\
& &{\hat{\mib C}}_2(m_2^{(2)},\cdots,m_M^{(2)})=
({\hat{\mib S}}_2^1)^{m_2^{(2)}}\cdots
({\hat{\mib S}}_M^1)^{m_M^{(2)}}\ . 
\label{6-13b}
\end{eqnarray}
\end{subequations}
The state $\ket{m(2)}$ is an eigenstate of 
$({\hat S}_i^i-{\hat S}_1^1)$ : 
\begin{equation}\label{6-14}
({\hat S}_i^i-{\hat S}_1^1)\ket{m(2)}
=-\left[(s_i-m_i^{(1)}-m_i^{(2)})
-(s_1-m_1^{(1)})-\sum_{k=2}^M m_k^{(2)}\right]\ket{m(2)}\ .
\end{equation}
The above means that the set $(\ket{m(2)})$ forms the 
orthogonal set. 
Thus, we finished the operation of ${\hat S}_i^1\ 
(i=2, \cdots, M)$ through ${\hat{\mib S}}_i^1$.

Next operation is related with the operator ${\hat S}_i^2\ 
(i=3,\cdots, M)$. 
In this case, we also note that the state $\ket{m(2)}$ 
is the intrinsic state for the $su(M-1)$ algebra. 
Therefore, we can apply the same idea as that already 
presented in the case of ${\hat S}_i^1$. 
By applying the above idea, successively, we arrive at the 
stage of the operation of ${\hat S}_i^n\ (i=n+1, \cdots, M)$. 
First, the set $({\hat S}_i^n, {\hat S}_n^i, \ (i=n+1, \cdots, M), 
{\hat S}_{n+1}^{n+1}-{\hat S}_n^n, {\hat S}_j^i-\delta_{ij}
{\hat S}_{n+1}^{n+1},\ (i, j=n+1, \cdots, M))$ forms the 
$su(M-n+1)$ algebra and the sub-set 
$({\hat S}_j^i-\delta_{ij}{\hat S}_{n+1}^{n+1})\ (i, j=n+1, \cdots, M)$ 
composes the $su(M-n)$ algebra. 
In this case, we also define the operator ${\hat{\mib S}}_i^n$ in the 
form 
\begin{equation}\label{6-15}
{\hat{\mib S}}_i^n=\sum_{l=n+1}^M{\hat S}_l^n{\hat U}_{il}^{(n)} \ .
\quad (i=n+1, \cdots, M)
\end{equation}
Here, ${\hat U}_{il}^{(n)}$ is a function of ${\hat S}_j^i\ 
(j>i=n+1, \cdots, M-1)$ and $({\hat S}_i^i-{\hat S}_{n+1}^{n+1})\ 
(i=n+2, \cdots, M)$. 
Of course, we regard ${\hat{\mib S}}_i^n$ as the operator 
satisfying the relations 
\begin{subequations}\label{6-16}
\begin{eqnarray}
& & [ {\hat S}_i^j , {\hat{\mib S}}_k^n ]
=\sum_{m>l=n+1,\cdots,M-1} {\hat V}_{ijk,lm}^{(n)}{\hat S}_l^m \ , 
\quad (j>i=n+1, \cdots, M-1) \qquad\label{6-16a}\\
& & [ {\hat S}_i^i-{\hat S}_n^n , {\hat{\mib S}}_k^n ]
=(1+\delta_{ik}){\hat{\mib S}}_k^n \ , 
\quad (i=n+1, \cdots, M) \label{6-16b}\\
& & [ {\hat{\mib S}}_i^n , {\hat{\mib S}}_j^n ]=0 \ .
\label{6-16c}
\end{eqnarray}
\end{subequations}
The relations (\ref{6-15}) and (\ref{6-16}) in the case $n=1$ 
are reduced to Eqs.(\ref{6-9}) and (\ref{6-10}). 
In the same way as that in the case $n=1$, the state 
$\ket{m(n+1)}$ is defined as 
\begin{subequations}\label{6-17}
\begin{eqnarray}
& &\ket{m(n+1)}={\hat{\mib C}}_{n+1}(m_{n+1}^{(n+1)}, \cdots,
m_{M}^{(n+1)}) \ket{m(n)} \ , \label{6-17a}\\
& &{\hat{\mib C}}_{n+1}(m_{n+1}^{(n+1)},\cdots, m_M^{(n+1)})
=({\hat{\mib S}}_{n+1}^n)^{m_{n+1}^{(n+1)}}\cdots
({\hat{\mib S}}_{M}^n)^{m_{M}^{(n+1)}} \ . \label{6-17b}
\end{eqnarray}
\end{subequations}
Under the successive application of the above idea, finally, 
we arrive at the case $n=M-1$. 
In this case, the set $({\hat S}_M^{M-1}, {\hat S}_{M-1}^M, 
{\hat S}_M^M-{\hat S}_{M-1}^{M-1})$ forms the 
$su(2)$ algebra and ${\hat{\mib S}}_i^n$ is reduced to 
${\hat{\mib S}}_M^{M-1}={\hat S}_M^{M-1}$.

By summarizing the above procedure, the state 
$\ket{m(M)}$ is expressed in the form 
\begin{eqnarray}\label{6-18}
\ket{m(M)}
&=&{\hat{\mib C}}_M(m_M^{(M)})
{\hat{\mib C}}_{M-1}(m_{M-1}^{(M-1)},m_M^{(M-1)})\nonumber\\
& &\times\cdots\times
{\hat{\mib C}}_{n+1}(m_{n+1}^{(n+1)},\cdots, m_M^{(n+1)})\nonumber\\
& &\times\cdots\times
{\hat{\mib C}}_{2}(m_{2}^{(2)},\cdots, m_M^{(2)})
{\hat{\mib C}}_{1}(m_{1}^{(1)},\cdots, m_M^{(1)})\ket{m}\ .
\end{eqnarray}
The above method can be applied to the case of the 
$su(M,1)$ algebra by replacing ${\hat S}^i$, ${\hat S}_i$, 
${\hat S}_j^i\ (j>i)$ and ${\hat S}_i^j\ (j>i)$ with 
${\hat T}^p$, ${\hat T}_p$, 
${\hat T}_p^q\ (q>p)$ and ${\hat T}_q^p\ (q>p)$, 
respectively. 
The state $\ket{m\mu(M)}$ in the $su(M+1)$ and in the $su(M,1)$ 
algebra can be expressed in the form 
\begin{eqnarray}\label{6-19}
\ket{m\mu(M)}
&=&{\hat{\mib D}}_M(\mu_M^{(M)})\cdots{\hat{\mib D}}_1(\mu_1^{(1)}, 
\cdots,\mu_M^{(1)}) \nonumber\\
& &\times
{\hat{\mib C}}_M(m_M^{(M)})\cdots
{\hat{\mib C}}_{1}(m_{1}^{(1)},\cdots, m_M^{(1)})
\ket{n, n_1, \cdots, n_M} \ . 
\end{eqnarray}
It may be not necessary to mention the meaning of 
${\hat{\mib D}}_M(\mu_M^{(M)})\cdots{\hat{\mib D}}_1(
\mu_1^{(1)}, \cdots, \mu_M^{(1)})$. 
Clearly, the state $\ket{m\mu(M)}$ contains $(M+1)^2$ 
quantum numbers and our problem is reduced to determine the 
operators ${\hat U}_{il}$ and ${\hat U}_{il}^{(n)}\ (n=1, 2, \cdots, 
M-1)$ appearing in Eqs.(\ref{6-1}) and (\ref{6-15}), respectively. 
Some examples were already given in Eqs.(\ref{6-3}), (\ref{6-4}), 
(\ref{6-5}), (\ref{6-11}) and (\ref{6-12}). 
Further, we should note that, in the form of $\ket{m}$ 
shown in Eq.(\ref{5-8}), ${\hat A}_M^*$ obeys the relation 
(\ref{4-14}). 
Therefore, the part $\ket{n, n_1, \cdots, n_M}$ shown in 
Eq.(\ref{6-19}) is replaced with 
\begin{equation}\label{6-20}
\ket{n, n_1, \cdots, n_M}
=({\hat A}_M^*)^{(N-S)/(M+1)-n_1}
\dket{s_1, s_2, \cdots, s_M} \ .
\end{equation}
Since we have the relation (\ref{4-14}), $\ket{m\mu(M)}$ 
can be rewritten in the form 
\begin{eqnarray}\label{6-21}
\ket{m\mu(M)}\ 
&=&({\hat A}_M^*)^{(N-S)/(M+1)-n_1}\dket{m\mu(M)} \ , \nonumber\\
\dket{m\mu(M)}
&=&{\hat{\mib D}}_M(\mu_M^{(M)})\cdots{\hat{\mib D}}_1(\mu_1^{(1)}, 
\cdots,\mu_M^{(1)}) \nonumber\\
& &\times
{\hat{\mib C}}_M(m_M^{(M)})\cdots
{\hat{\mib C}}_{1}(m_{1}^{(1)},\cdots, m_M^{(1)})
\dket{s_1, s_2, \cdots, s_M} \ . 
\end{eqnarray}
The role of the part related with ${\hat A}_M^*$ in the state 
$\ket{m\mu(M)}$ can be interpreted in a way mentioned below. 
We note the relation 
\begin{equation}\label{6-22}
[ {\hat A}_M^* , {\hat S} ]=(M+1){\hat A}_M^* \ .
\end{equation}
As was shown in the inequality (\ref{5-5}), $S$ runs 
in the region between $S_{\rm max}$ and $S_{\rm min}$ : 
\begin{equation}\label{6-23}
S_{\rm max}=N-(M+1)n_1 \ , \qquad
S_{\rm min}=N-(M+1)(\Omega-n) \ .
\end{equation}
If $S=S_{\rm max}$, $(N-S)/(M+1)-n_1=0$ and 
$\ket{m\mu(M)}=\dket{m\mu(M)}$. 
By $\rho$-time operation of ${\hat A}_M^*$ on 
$\dket{m\mu(M)}$, we have the state 
with $S=S_{\rm max}-(M+1)\rho$ and, finally, the 
$s_1$-time operation gives us the state 
with $S=S_{\rm min}$. 
Therefore, ${\hat A}_M^*$ plays a role of the 
lowering operator for ${\hat S}$. 
For obtaining the representation of the $su(M+1)$ and 
the $su(M,1)$ algebra, it is enough to take into account the 
state $\dket{m\mu(M)}$.

\section{Discussion and concluding remarks}

In this section, mainly, we will discuss two concrete examples, 
the $su(2)$ and the $su(3)$ algebra. 
Let us start from the case of the $su(2)$ algebra. 
This case corresponds to $M=N=1$ and the $su(2)$ and the 
$su(1,1)$ generators are expressed in the following form : 
\begin{subequations}\label{7-1}
\begin{eqnarray}
& &{\hat S}^1={\hat a}_1^*{\hat b}+{\hat a}^{1*}{\hat b}_1^1 \ , 
\qquad
{\hat S}^2={\hat b}^*{\hat a}_1+{\hat b}_1^{1*}{\hat a}^1 \ , 
\nonumber\\
& &{\hat S}_1^1={\hat a}_1^*{\hat a}_1-
{\hat b}_1^{1*}{\hat b}_1^1+{\hat a}^{1*}{\hat a}^1-
{\hat b}^*{\hat b} \ , \label{7-1a}\\
& &{\hat T}^1={\hat a}_1^*{\hat b}^*
-{\hat a}_1^{*}{\hat b}_1^{1*} \ , 
\qquad
{\hat T}^2={\hat b}{\hat a}_1-{\hat b}_1^{1}{\hat a}_1 \ , 
\nonumber\\
& &{\hat T}_1^1={\hat a}^{1*}{\hat a}^1+
{\hat b}_1^{1*}{\hat b}_1^1+{\hat a}_1^{*}{\hat a}_1+
{\hat b}^*{\hat b}+2 \ . \label{7-1b}
\end{eqnarray}
\end{subequations}
The above algebras are constructed in terms of four kinds 
of the boson operators $({\hat a}_1, {\hat a}_1^*)$, 
$({\hat b}, {\hat b}^*)$, $({\hat a}^1, {\hat a}^{1*})$ 
and $({\hat b}_1^1, {\hat b}_1^{1*})$ and 
the form (\ref{7-1a}) should be compared with the form (\ref{1-13}). 
The operator ${\hat S}$ is expressed as 
\begin{equation}\label{7-2}
{\hat S}={\hat a}_1^*{\hat a}_1-
{\hat a}^{1*}{\hat a}^1-{\hat b}_1^{1*}{\hat b}_1^1+
{\hat b}^*{\hat b} \ .
\end{equation}
Further, ${\hat A}_1^*$ is given in the form 
\begin{equation}\label{7-3}
{\hat A}_1^*={\hat B}_1^*{\hat b}+
{\hat B}_1^{*}(1){\hat a}_1
={\hat b}_1^{1*}{\hat b}+{\hat a}^{1*}{\hat a}_1 \ .
\end{equation}

Since we treat the case $M=1$, ${\hat{\mib D}}_1(\mu_1^{(1)})
=({\hat T}^1)^{\mu_1^{(1)}}$ and 
${\hat{\mib C}}_1(m_1^{(1)})=({\hat S}^1)^{m_1^{(1)}}$. 
Therefore, $\ket{m\mu(1)}$ can be expressed in the following form : 
\begin{equation}\label{7-4}
\ket{m\mu(1)}=({\hat A}_1^*)^{(s_1-S)/2}
({\hat T}^1)^{(\mu-(s_1+2))/2}
({\hat S}^1)^{(s_1+m)/2}({\hat b}^*)^{s_1}\ket{0} \ .
\end{equation}
Here, 
\begin{eqnarray}\label{7-5}
& &(N-S)/2-n_1=(s_1-S)/2 \ , \nonumber\\
& &\mu_1^{(1)}=(\mu-(s_1+2))/2 \ , \qquad m_1^{(1)}=(s_1+m)/2 \ , 
\nonumber\\
& &s_1=\Omega-n-n_1 \ .
\end{eqnarray}
The state $\ket{m\mu(1)}$ is the eigenstate of 
${\hat S}$, ${\hat \Gamma}_{su(1,1)}$, ${\hat T}_1^1$, 
${\hat \Gamma}_{su(2)}$ and ${\hat S}_1^1$ with the eigenvalues $S$, 
$(1/2)(s_1+2)((s_1+2)-2)$, $\mu$, 
$(1/2)s_1(s_1+2)$ and $m$, respectively. 
Of course, $\mu=s_1+2, s_1+4, s_1+6, \cdots$ and 
$m=-s_1, -s_1+2, \cdots, s_1-2, s_1$. 
Since $S_{\rm min}=N-2(\Omega-n)=-s_1$ and 
$S_{\rm max}=N-2n_1=s_1$, we have 
\begin{equation}\label{7-6}
S=-s_1, -s_1+2, \cdots, s_1-2, s_1 \ .
\end{equation}

Concerning with the relation (\ref{7-6}), we must give an interesting 
remark. 
The following set obeys the $su(2)$ algebra : 
\begin{equation}\label{7-7}
{\hat R}^1={\hat A}_1^* \ , \qquad
{\hat R}_1={\hat A}_1\ , \qquad 
{\hat R}_1^1=-{\hat S} \ .
\end{equation}
Of course, the set $({\hat R}^1, {\hat R}_1, {\hat R}_1^1)$ 
commutes with $({\hat S}^1, {\hat S}_1, {\hat S}_1^1)$ 
and $({\hat T}^1, {\hat T}_1, {\hat T}_1^1)$ and the 
state $\ket{m\mu(1)}$ is the eigenstate 
of the Casimir operator ${\hat \Gamma}_{su(2)}'\ 
(={\hat R}^1{\hat R}_1+{\hat R}_1{\hat R}^1+(1/2)({\hat R}_1^1)^2)$ 
with the eigenvalue $(1/2)s_1(s_1+2)$. 
The details of the above three algebras have been already discussed by 
three of the present authors (A. K., J. P. \& M. Y.).\cite{KPY2}

Our next interest is concerned with the case $M=N=2$. 
The basic operators can be expressed in the following form :
\begin{subequations}\label{7-8}
\begin{eqnarray}
& &{\hat S}^1={\hat a}_1^*{\hat b}+{\hat a}^{1*}{\hat b}_1^1
+{\hat a}^{2*}{\hat b}_1^2 \ , \qquad
{\hat S}_1={\hat b}^*{\hat a}_1+{\hat b}_1^{1*}{\hat a}^1
+{\hat b}_1^{2*}{\hat a}^2 \ , \nonumber\\
& &{\hat S}^2={\hat a}_2^*{\hat b}+{\hat a}^{1*}{\hat b}_2^1
+{\hat a}^{2*}{\hat b}_2^2 \ , \qquad
{\hat S}_2={\hat b}^*{\hat a}_2+{\hat b}_2^{1*}{\hat a}^1
+{\hat b}_2^{2*}{\hat a}^2 \ , \nonumber\\
& &{\hat S}_2^1={\hat a}_2^*{\hat a}_1-{\hat b}_1^{1*}{\hat b}_2^1
-{\hat b}_1^{2*}{\hat b}_2^2 \ , \qquad
{\hat S}_1^2={\hat a}_1^*{\hat a}_2-{\hat b}_2^{1*}{\hat b}_1^1
-{\hat b}_2^{2*}{\hat b}_1^2 \ , \nonumber\\
& &{\hat S}_1^1={\hat a}_1^*{\hat a}_1-{\hat b}_1^{1*}{\hat b}_1^1
-{\hat b}_1^{2*}{\hat b}_1^2+{\hat a}^{1*}{\hat a}^1
+{\hat a}^{2*}{\hat a}^2-{\hat b}^{*}{\hat b} \ , \nonumber\\
& &{\hat S}_2^2={\hat a}_2^*{\hat a}_2-{\hat b}_2^{1*}{\hat b}_2^1
-{\hat b}_2^{2*}{\hat b}_2^2+{\hat a}^{1*}{\hat a}^1
+{\hat a}^{2*}{\hat a}^2-{\hat b}^{*}{\hat b} \ , \label{7-8a}\\
& &{\hat T}^1={\hat a}^{1*}{\hat b}^*-{\hat a}_1^{*}{\hat b}_1^{1*}
-{\hat a}_2^{*}{\hat b}_2^{1*} \ , \qquad
{\hat T}_1={\hat b}{\hat a}^1-{\hat b}_1^{1}{\hat a}_1
-{\hat b}_2^{1}{\hat a}_2 \ , \nonumber\\
& &{\hat T}^2={\hat a}^{2*}{\hat b}^*-{\hat a}_1^{*}{\hat b}_1^{2*}
-{\hat a}_2^{*}{\hat b}_2^{2*} \ , \qquad
{\hat T}_2={\hat b}{\hat a}^2-{\hat b}_1^{2}{\hat a}_1
-{\hat b}_2^{2}{\hat a}_2 \ , \nonumber\\
& &{\hat T}_1^2={\hat a}^{2*}{\hat a}^1+{\hat b}_1^{2*}{\hat b}_1^1
+{\hat b}_2^{2*}{\hat b}_2^1 \ , \qquad
{\hat T}_2^1={\hat a}^{1*}{\hat a}^2+{\hat b}_1^{1*}{\hat b}_1^2
+{\hat b}_2^{1*}{\hat b}_2^2 \ , \qquad\quad
\nonumber\\
& &{\hat T}_1^1={\hat a}^{1*}{\hat a}^1+{\hat b}_1^{1*}{\hat b}_1^1
+{\hat b}_2^{1*}{\hat b}_2^1+{\hat a}_1^{*}{\hat a}_1
+{\hat a}_2^{*}{\hat a}_2+{\hat b}^{*}{\hat b}+3 \ , \nonumber\\
& &{\hat T}_2^2={\hat a}^{2*}{\hat a}^2+{\hat b}_1^{2*}{\hat b}_1^2
+{\hat b}_2^{2*}{\hat b}_2^2+{\hat a}_1^{*}{\hat a}_1
+{\hat a}_2^{*}{\hat a}_2+{\hat b}^{*}{\hat b}+3 \ , \label{7-8b}
\end{eqnarray}
\end{subequations}
\vspace{-0.8cm}
\begin{eqnarray}
& &{\hat S}={\hat a}_1^*{\hat a}_1+{\hat a}_2^*{\hat a}_2
-{\hat a}^{1*}{\hat a}^1-{\hat a}^{2*}{\hat a}^2 
-{\hat b}_1^{1*}{\hat b}_1^1-{\hat b}_2^{1*}{\hat b}_2^1
-{\hat b}_1^{2*}{\hat b}_1^2-{\hat b}_2^{2*}{\hat b}_2^2 \ , 
\label{7-9}\\
& &{\hat A}_2^*
=({\hat b}_2^{1*}{\hat b}_1^{2*}-{\hat b}_1^{1*}{\hat b}_2^{2*})
{\hat b}
+({\hat b}_1^{2*}{\hat a}^{1*}-{\hat b}_1^{1*}{\hat a}^{2*})
{\hat a}_1
+({\hat b}_2^{1*}{\hat a}^{2*}-{\hat b}_2^{2*}{\hat a}^{1*})
{\hat a}_2  \ , 
\label{7-10}\\
& &{\hat{\mib S}}^1={\hat S}^1\ , \qquad
{\hat{\mib S}}^2={\hat S}^1\cdot{\hat S}_2^1+{\hat S}^2\cdot
({\hat S}_2^2-{\hat S}_1^1) \ , \qquad
{\hat{\mib S}}_2^1={\hat S}_2^1 \ , \label{7-11}\\
& &{\hat{\mib T}}^1={\hat T}^1\ , \qquad
{\hat{\mib T}}^2={\hat T}^1\cdot{\hat T}_1^2+{\hat T}^2\cdot
({\hat T}_2^2-{\hat T}_1^1) \ , \qquad
{\hat{\mib T}}_1^2={\hat T}_1^2 \ . \label{7-12}
\end{eqnarray}

With the use of the above basic operators, let us construct 
the orthogonal set for the present case. 
For this aim, the quantum numbers shown in the relations 
(\ref{5-9}) are convenient : 
\begin{eqnarray}
& &s^1=n_1-n_2 \ , \qquad
s^2=\Omega-n-(n_1+n_2)/2 \ , \label{7-13}\\
& &t^1=s^1\ , \qquad\quad t^2=s^2+3 \ . \label{7-14}
\end{eqnarray}
Inversely, $n_1$ and $n_2$ are expressed as 
\begin{equation}\label{7-13a}
n_1=\Omega-n+s^1/2-s^2 \ , \qquad
n_2=\Omega-n-s^1/2-s^2 \ .
\end{equation}
The eigenvalues of the $su(3)$ and the $su(2,1)$ algebra 
are given in the form 
\begin{eqnarray}\label{7-15}
& &\gamma_{su(3)}^{(b)}=(1/2)s^1(s^1+2)
+(2/3)s^2(s^2+3) \ , \nonumber\\
& &\gamma_{su(2,1)}^{(b)}=(1/2)s^1(s^1+2)
+(2/3)(s^2+3)[(s^2+3)-3] \  \nonumber\\
\end{eqnarray}
Since $n_1-n_2=s^1$ and $\Omega-n-n_1=s^2-s^1/2$, 
$\dket{s_1,s_2}$ is expressed as 
\begin{equation}\label{7-16}
\dket{s_1,s_2}=({\hat b}_2^{1*})^{s^1}
({\hat b}^*)^{(2s^2-s^1)/2}\ket{0} \ .
\end{equation}
Further, the quantity $(N-S)/3-n_1$ is given as 
\begin{equation}\label{7-17}
(N-S)/3-n_1=(s^2-S)/3-s^1/2 \ ,
\end{equation}
\vspace{-0.8cm}
\begin{subequations}\label{7-18}
\begin{eqnarray}
& &S_{\rm min}=N-3(\Omega-n)=-2s^2 \ , \label{7-18a}\\
& &S_{\rm max}=N-3n_1=3(s^2/3-s^1/2) \ .\label{7-18b}
\end{eqnarray}
\end{subequations}
Therefore, we have 
\begin{equation}\label{7-19}
0\le (N-S)/3-n_1 \le (2s^2-s^1)/2 \ .
\end{equation}
The operator operated on the intrinsic state can be expressed in 
the form 
\begin{eqnarray}\label{7-20}
& &{\hat{\mib D}}_2(\mu_2^{(2)}){\hat{\mib D}}_1(\mu_1^{(1)}, 
\mu_2^{(1)})
{\hat{\mib C}}_2(m_2^{(2)}){\hat{\mib C}}_1(m_1^{(1)}, m_2^{(1)})
\nonumber\\
&=&({\hat{\mib T}}_1^2)^{\mu_2^{(2)}}({\hat{\mib T}}^1)^{\mu_1^{(1)}}
({\hat{\mib T}}^2)^{\mu_2^{(1)}}({\hat{\mib S}}_2^1)^{m_2^{(2)}}
({\hat{\mib S}}^1)^{m_1^{(1)}}({\hat{\mib S}}^2)^{m_2^{(1)}} \ .
\end{eqnarray}
Therefore, the orthogonal set is obtained in the following form :
\begin{eqnarray}\label{7-21}
\ket{m\mu(2)}
&=&({\hat A}_2^*)^{(s^2-S)/3-s^1/2}\nonumber\\
& &\times{\hat{\mib D}}_2(\mu_2^{(2)}){\hat{\mib D}}_1(\mu_1^{(1)}, 
\mu_2^{(1)})
{\hat{\mib C}}_2(m_2^{(2)}){\hat{\mib C}}_1(m_1^{(1)}, m_2^{(1)})
\nonumber\\
& &\times
({\hat b}_2^{1*})^{s^1}({\hat b}^*)^{(2s^2-s^1)/2}\ket{0} \ .
\end{eqnarray}

If $S=S_{\rm max}$ and $\mu_2^{(2)}=\mu_1^{(1)}=\mu_2^{(1)}=0$, 
the state (\ref{7-21}) is expressed as 
\begin{equation}\label{7-22}
\ket{m\mu(2)}=({\hat{\mib S}}_2^1)^{m_2^{(2)}}
({\hat{\mib S}}^1)^{m_1^{(1)}}({\hat{\mib S}}^2)^{m_2^{(1)}}
({\hat b}_2^{1*})^{s^1}({\hat b}^*)^{(2s^2-s^1)/2}\ket{0} \ .
\end{equation}
The set (\ref{7-22}) is the orthogonal set for the irreducible 
representation of the $su(2)$ algebra. 
Concerning the state (\ref{7-22}), we will give a remark. 
The set $({\hat S}_2^1,{\hat S}_1^2, {\hat S}_2^2-{\hat S}_1^1)$ 
forms the $su(2)$ algebra and we adopt the conventional 
notation for the $su(2)$ algebra : 
\begin{equation}\label{7-23}
{\hat J}_+={\hat S}_2^1 \ , \qquad
{\hat J}_-={\hat S}_1^2\ , \qquad
{\hat J}_0=({\hat S}_2^2-{\hat S}_1^1)/2 \ .
\end{equation}
Further, we can prove that ${\hat S}^1$ and ${\hat S}^2$ play a role 
of the spherical tensor with rank $1/2$ and we denote 
them as 
\begin{equation}\label{7-24}
{\hat C}_{-1/2}^*={\hat S}^1 \ , \qquad
{\hat C}_{+1/2}^*={\hat S}^2 \ .
\end{equation}
Then, ${\hat{\mib S}}^1$ and ${\hat{\mib S}}^2$ are expressed as 
\begin{eqnarray}\label{7-25}
& &{\hat{\mib S}}^1={\hat C}_{-1/2}^*={\hat{\mib C}}_{-1/2}^* \ , \nonumber\\
& &{\hat{\mib S}}^2={\hat C}_{-1/2}^*\cdot{\hat J}_+
+2{\hat C}_{+1/2}^*\cdot{\hat J}_0
=2{\hat{\mib C}}_{+1/2}^* \ .
\end{eqnarray}
With the use of the above notations, the state (\ref{7-22}) 
is rewritten in the form (except the normalization) 
\begin{subequations}\label{7-230}
\begin{equation}\label{7-23a}
\ket{m\mu(2)}=({\hat J}_+)^{j_3+m_3}
({\hat{\mib C}}_{-1/2}^*)^{-j_3+j_2+j_1}
({\hat{\mib C}}_{+1/2}^*)^{j_3+j_2-j_1}
\dket{j_1, s^2} \ .
\end{equation}
Here, we put 
\begin{equation}\label{7-23b}
m_1^{(1)}=-j_3+j_2+j_1 \ , \qquad
m_2^{(1)}=j_3+j_2-j_1\ , \qquad
s^1=2j_1\ .
\end{equation}
\end{subequations}
The state $\dket{j_1, s^2}$ denotes 
\begin{equation}\label{7-240}
\dket{j_1, s^2}=({\hat b}_2^{1*})^{s^1}({\hat b}^*)^{(2s^2-s^1)/2}
\ket{0} \ .
\end{equation}
The quantities $j_1$, $j_2$ and $j_3$ denote 
\begin{equation}\label{7-250}
j_1,\ j_2,\ j_3=0,\ 1/2,\ 1,\ 3/2,\ \cdots .
\end{equation}
We denote the minimum weight state $\dket{j, s^2}$ for the 
spin $j$ : 
\begin{equation}\label{7-26}
{\hat J}_-\dket{j, s^2}=0 \ , \qquad
{\hat J}_0\dket{j, s^2}=-j\dket{j, s^2}\ .
\end{equation}
The state $\dket{j_1, s^2}$ shown in Eq.(\ref{7-240}) 
is an example of $\dket{j, s^2}$. 
Except the normalization constant, the state $\dket{j, s^2}$ 
connects with $\dket{j\pm1/2, s^2}$ through 
\begin{equation}\label{7-27}
\dket{j, s^2}={\hat{\mib C}}_{\mp1/2}^* \dket{j\pm1/2, s^2} \ .
\end{equation}
For the prove of the relation (\ref{7-27}), we use the relations 
\begin{equation}\label{7-28}
[ {\hat J}_- , {\hat{\mib C}}_{\mp1/2}^* ]
={\hat {\mib C}}_{\pm 1/2}^* {\hat J}_- \ , \qquad
[ {\hat J}_0 , {\hat{\mib C}}_{\mp1/2}^* ]
=\pm(1/2){\hat{\mib C}}_{\mp 1/2}^* \ . 
\end{equation}
The relation (\ref{7-27}) tells us that the operators 
${\hat{\mib C}}_{\mp 1/2}^*$ play a role of the lowering 
(the upper sign) and the raising (the lower sign) operator for 
the state $\dket{j, s^2}$. 
With the successive use of the relation (\ref{7-27}), 
we obtain 
\begin{equation}\label{7-29}
\dket{j_3, s^2}
=({\hat{\mib C}}_{-1/2}^*)^{-j_3+j_2+j_1}
({\hat{\mib C}}_{+1/2}^*)^{j_3+j_2-j_1}\dket{j_1, s^2} \ .
\end{equation}
Therefore, $\ket{m\mu(2)}$ shown in Eq.(\ref{7-23}) 
can be written as 
\begin{equation}\label{7-30}
\ket{m\mu(2)}=\ket{j_1j_2; j_3 m_3, s^2} \ .
\end{equation}
The state (\ref{7-29}) can be rewritten in the form 
\begin{eqnarray}\label{7-31}
\dket{j_3, s^2}&=&
({\hat{\mib C}}_{+1/2}^*)^{j_3+j_2-j_1}
({\hat{\mib C}}_{-1/2}^*)^{-j_3+j_2+j_1}\dket{j_1, s^2} \nonumber\\
&=&({\hat{\mib C}}_{+1/2}^*)^{j_3+j_2-j_1} 
\dket{(j_3-j_2+j_1)/2 , s^2} \ .
\end{eqnarray}
Since $-j_3+j_2+j_1 \ge 0$, $j_3+j_2-j_1 \ge 0$ and 
$j_3-j_2+j_1 \ge 0$, we have 
\begin{equation}\label{7-32}
|j_1-j_2|\le j_3 \le j_1+j_2 \ .
\end{equation}
We can see that the state (\ref{7-30}) can be regarded as the state 
\begin{equation}\label{7-33}
\ket{j_1j_2; j_3 m_3, s^2}
=\sum_{m_1,m_2}\bra{j_1m_1j_2m_2}j_3m_3\rangle 
{\hat{\mib C}}_{j_2m_2}^*\ket{j_1m_1, s^2} \ .
\end{equation}
Here, $\bra{j_1m_1j_2m_2}j_3m_3\rangle$ denotes the 
Clebsch-Gordan coefficient and ${\hat{\mib C}}_{j_2m_2}^*$ 
is defined as 
\begin{equation}\label{7-39}
{\hat{\mib C}}_{j_2m_2}^*
=\left(\sqrt{(j_2+m_2)!(j_2-m_2)!}\right)^{-1}
({\hat{\mib C}}_{+1/2}^*)^{j_2+m_2}({\hat{\mib C}}_{-1/2}^*)^{j_2-m_2} 
\ .
\end{equation}
The operator ${\hat{\mib C}}_{j_2m_2}^*$ denotes spherical tensor with 
rank $j_2$. 
The state $\ket{j_1m_1, s^2}$ is given as 
\begin{equation}\label{7-40}
\ket{j_1m_1, s^2}=({\hat J}_+)^{j_1+m_1}
\dket{j_1, s^2} \ .
\end{equation}

Finally, we will give short concluding remarks. 
In this paper, we developed the generalized Lipkin model in 
the Schwinger boson representation. 
Not only the symmetric representation but also non-symmetric representation 
for the $su(M+1)$ algebra was formulated. 
At the same time, the $su(M,1)$ algebra was also treated. 
We already knew that the non-compact algebra, for example, 
the $su(1,1)$ algebra, plays an interesting role for describing 
various thermal phenomenon, such as non-equilibrium state.\cite{KPTY} 
We can find the reason why such descriptions are possible 
in the phase space doubling. 
The $su(M,1)$ algebra may play the role of the phase space doubling. 
This is our next problem.

\section*{Acknowledgements}

Main part of this work was performed when two of the present authors 
(Y. T. \& M. Y.) stayed at Coimbra in September of 2000. 
They express their sincere thanks to Professor 
J. da Provid\^encia, co-author of this paper, for his kind invitation 
to Coimbra. This work was partially supported by the 
Kansai University Grant-in-Aid for the Faculty Joint 
Research Program, 2000.

\appendix
\section{The eigenvalues of the Casimir operators} 

First, we investigate the quadratic form $\sum_{c,d=1}^L
\Lambda_{cd}X_c X_d$. 
Here, $X_c$, $X_d$ and $\Lambda_{cd}$ are real and 
$\Lambda_{cd}=\Lambda_{dc}$. 
The orthogonal transformation permits us to express the 
above quadratic form in the following form : 
\begin{equation}\label{A-1}
\sum_{c,d=1}^L\Lambda_{cd}X_c X_d = \sum_{e=1}^L\lambda_e(X'_e)^2 \ .
\end{equation}
The quantity $\lambda_e$ denotes the eigenvalue of the symmetric 
matrix $(\Lambda_{cd})$. 
The relation between two vectors $(X_c)$ and $(X'_e)$ are given by 
\begin{equation}\label{A-2}
X_e'=\sum_{c=1}^L y_c^{(e)}X_c \ , \qquad
X_c=\sum_{e=1}^L y_c^{(e)}X_e' \ .
\end{equation}
Here, $(y_c^{(e)})$ is an orthogonal matrix obeying the 
conditions 
\begin{eqnarray}
& &\sum_{d=1}^L\Lambda_{cd}y_d^{(e)}
=\lambda_e y_c^{(e)} \ , \quad (e=1,2,\cdots, L) \label{A-3}\\
& &\sum_{c=1}^L y_c^{(e)}y_c^{(f)}=\delta_{ef} \ , \qquad
\sum_{e=1}^L y_c^{(e)}y_d^{(e)} = \delta_{cd} \ . \label{A-4}
\end{eqnarray}
In this Appendix, we will treat the following form : 
\begin{equation}\label{A-5}
\Lambda_{cd}=\delta_{cd}-(L+1)^{-1} \ .
\end{equation}

It can be easily shown that the eigenvalue equation (\ref{A-3}) 
for the form (\ref{A-5}) has the following solution : 
\begin{equation}\label{A-6}
\lambda_e=\cases{
1 \ , \qquad (e=1,2,\cdots,L-1) \cr
(L+1)^{-1} \ . \quad (e=L)}
\end{equation}
For the case $e=L$, $y_c^{(L)}$ is given by 
\begin{equation}\label{A-7}
y_c^{(L)}=1/\sqrt{L} \ . \quad (c=1,2,\cdots, L)
\end{equation}
However, for the case $e=1,2,\cdots, L-1$, all $\lambda_e$ are 
equal to 1 and, then, it is impossible to determine $y_c^{(e)}$ 
uniquely. 
Only we can show that $y_c^{(e)}$ should obey 
\begin{eqnarray}\label{A-8}
& &\sum_{c=1}^L y_c^{(e)}=0 \ , \qquad\qquad (e=1,2,\cdots, L-1) 
\nonumber\\
& &\sum_{c=1}^L y_c^{(e)}y_c^{(f)}=\delta_{ef} \ . 
\quad
(e, f= 1,2,\cdots, L-1)
\end{eqnarray}
With the use of the above result, the relation (\ref{A-1}) becomes 
of the form 
\begin{equation}\label{A-9}
\sum_{c=1}^L(X_c)^2-(L+1)^{-1}\left(\sum_{c=1}^L X_c\right)^2
=\sum_{e=1}^{L-1}(X_e')^2+(L+1)^{-1}(X_L')^2 \ .
\end{equation}

For the convenience of the discussion in \S\S 3 and 4, we 
adopt the following form for $y_c^{(e)}$ in $e=1,2,\cdots,L-1$ : 
\begin{equation}\label{A-10}
y_c^{(e)}=\cases{
-(\sqrt{e(e+1)})^{-1} \ , \quad (c=1,2,\cdots,e) \cr
\sqrt{e/(e+1)} \ , \quad\ \ (c=e+1) \cr
0 \ , \qquad\quad (c=e+2,\cdots, L)}
\end{equation}
or
\begin{equation}\label{A-11}
y_c^{(e)}=\cases{
0 \ , \qquad\quad (e=1,2,\cdots, c-2)\cr
\sqrt{e/(e+1)} \ , \quad\ \ (e=c-1) \cr
-(\sqrt{e(e+1)})^{-1} \ . \quad (e=c, \cdots,L-1)
}
\end{equation}
Direct calculation tells us that the above $y_c^{(e)}$ 
satisfies the relations (\ref{A-8}) and the relation 
\begin{equation}\label{A-12}
2\sum_{c=1}^L cy_c^{(e)}=\sqrt{e(e+1)} \ . \quad
(e=1, 2, \cdots, L-1)
\end{equation}
The relation (\ref{A-12}) leads us to the following form : 
\begin{equation}\label{A-13}
\sum_{c=1}^L (L-2c)X_c 
=-\sum_{e=1}^{L-1}\sqrt{e(e+1)}X_e'
-\sqrt{L}X_L' \ .
\end{equation}
The form (\ref{A-13}) is used in \S\S 3 and 4. 
Thus, we have
\begin{eqnarray}\label{A-14}
& &\sum_{c=1}^L(X_c)^2-(L+1)^{-1}\left(\sum_{c=1}^L X_c\right)^2
\mp\sum_{c=1}^L (L-2c)X_c \nonumber\\
& &\ \ =
\sum_{e=1}^{L-1}
[(X_e')^2\pm\sqrt{e(e+1)}X_e']
+(L+1)^{-1}(X_L')^2\pm\sqrt{L}X_L' \ .
\end{eqnarray}
For the convenience of the discussion in \S\S 3 and 4, we 
introduce new parameter $X^e\ (e=1, 2, \cdots, L)$ 
as follows :
\begin{equation}\label{A-15}
X^e=\sqrt{(e+1)/e}\ X_e' \ , \quad
(e=1, 2, \cdots, L-1) \ , 
\qquad
X^L=(1/\sqrt{L})X_L' \ .
\end{equation}
Then, the relation (\ref{A-14}) can be rewritten in the form 
\begin{eqnarray}
& &\sum_{c=1}^L(X_c)^2-(L+1)^{-1}\left(\sum_{c=1}^L X_c\right)^2
\mp\sum_{c=1}^L (L-2c)X_c\nonumber\\
& &\ \ =\sum_{e=1}^{L-1}[e/(e+1)]X^e[X^e\pm(e+1)]
+[L/(L+1)]X^L[X^L\pm(L+1)] \ , \quad\label{A-16}\\
& &X^e=\cases{-(1/e){\displaystyle \sum_{c=1}^e} X_c+X_{e+1} \ , \quad
(e=1,2,\cdots, L-1) \cr
(1/L){\displaystyle \sum_{c=1}^L}X_c \ , \qquad\qquad (e=L)}
\label{A-17}
\end{eqnarray}
or 
\begin{equation}\label{A-18}
X_c=[(c-1)/c]X^{c-1}-\sum_{e=c}^{L-1}(e+1)^{-1}X^e+X^L \ .
\end{equation}

\section{The proof of the properties of ${\hat B}_r^*$}

Let us give the proof of the properties of ${\hat B}_r^*$, 
which is defined in the form (\ref{4-4}). 
First, we note the following relations : 
\begin{subequations}\label{B-1}
\begin{eqnarray}
& &[ {\hat S}_i^j , {\hat b}_i^{s*} ] = -{\hat b}_j^{s*} \ , 
\label{B-1a}\\
& &[ {\hat S}_i^j , {\hat b}_k^{s*} ] = 0 \ , 
\quad (k\neq i) \label{B-1b}
\end{eqnarray}
\end{subequations}
\vspace{-0.9cm}
\begin{subequations}\label{B-2}
\begin{eqnarray}
& &[ {\hat T}_q^p , {\hat b}_k^{q*} ] = {\hat b}_k^{p*} \ , 
\label{B-2a}\\
& &[ {\hat T}_q^p , {\hat b}_k^{s*} ] = 0 \ . 
\quad (s\neq q) \label{B-2b}
\end{eqnarray}
\end{subequations}
It should be noted that by calculating the commutation relations 
(\ref{B-1a}) and (\ref{B-2a}), we obtain ${\hat b}_j^{s*}$ 
and ${\hat b}_k^{p*}$ from ${\hat b}_i^{s*}$ and ${\hat b}_k^{q*}$ 
by changing $i$ and $q$ with $j$ and $p$, respectively. 
For the case $i=1, 2, \cdots, M-r$, ${\hat B}_r^*$ 
does not contain ${\hat b}_i^{s*}$ and, then, in this case, we have 
the relation (\ref{4-6a}). 
For the case $i=M-r+1, \cdots, M$, we make the cofactor expansion 
for the determinant ${\hat B}_r^*$ : 
\begin{equation}\label{B-3}
{\hat B}_r^*=\sum_{s=1}^r {\hat b}_i^{s*} {\hat \Delta}_s^{i*} \ .
\end{equation}
Here, ${\hat \Delta}_s^{i*}$ denotes the cofactor of $(s, i)$. 
Then, the commutation relation $[{\hat S}_i^j , {\hat B}_r^* ]$ 
for $(j>i)$ gives us the following relation : 
\begin{eqnarray}\label{B-4}
[ {\hat S}_i^j , {\hat B}_r^* ]
&=&-\sum_{s=1}^r {\hat b}_j^{s*}{\hat \Delta}_s^{i*} \nonumber\\
&=&-\left|
\matrix{ {\hat b}_M^{1*} & \cdots & {\hat b}_M^{r*} \cr
          \cdots & \cdots & \cdots \cr
         {\hat b}_j^{1*} & \cdots & {\hat b}_j^{r*} \cr
          \cdots & \cdots & \cdots \cr
         {\hat b}_j^{1*} & \cdots & {\hat b}_j^{r*} \cr
          \cdots & \cdots & \cdots \cr
         {\hat b}_{M-r+1}^{1*} & \cdots & {\hat b}_{M-r+1}^{r*} 
         }
         \right|
=0 \ .
\end{eqnarray}
The reason is as follows : 
For the sake of the relation (\ref{B-1a}), 
${\hat b}_i^{s*}$ in ${\hat B}_r^*$ changes to ${\hat b}_j^{s*}$, 
which exists already in the $j$-th row $(j>i)$. 
Therefore, the determinant becomes vanish. 
Thus, we have the relation (\ref{4-6a}). 
The case (\ref{4-6b}) is also in the same situation as the above. 
First, we make the cofactor expansion : 
\begin{equation}\label{B-5}
{\hat B}_r^* = \sum_{k=M-r+1}^M {\hat b}_k^{q*}
{\hat \Delta}_q^{k*} \ .
\end{equation}
If $q=1, 2, \cdots, r$ and $q>p$, we have 
\begin{eqnarray}\label{B-6}
[ {\hat T}_q^p , {\hat B}_r^* ]
&=&\sum_{k=M-r+1}^M {\hat b}_k^{p*}{\hat \Delta}_q^{k*} \nonumber\\
&=&\left|
\matrix{ {\hat b}_M^{1*} & \vdots & {\hat b}_M^{p*} & \vdots & 
           {\hat b}_M^{p*} & \vdots & {\hat b}_M^{r*} \cr
         \vdots & \vdots& \vdots& \vdots& \vdots& \vdots& \vdots \cr
         {\hat b}_{M-r+1}^{1*} & \vdots & {\hat b}_{M-r+1}^{p*} & 
           \vdots & {\hat b}_{M-r+1}^{p*} & \vdots & {\hat b}_{M-r+1}^{r*} 
         }
         \right|
=0 \ .
\end{eqnarray}
The reason why the determinant vanishes is the same as the 
previous case. 
In the case $q=r+1, \cdots, M$, ${\hat B}_r^*$ does not 
contain ${\hat b}_k^{q*}$ in ${\hat B}_r^*$ and, then, 
the commutation relation $[ {\hat T}_q^p , {\hat B}_r^* ]$ 
becomes vanish. 
Thus, we obtain the relation (\ref{4-6b}).

Next, let us prove the relations (\ref{4-9a}) and (\ref{4-9b}). 
In the case $i=1, 2, \cdots, M-r$, ${\hat B}_r^*$ does not 
contain ${\hat b}_i^{s*}$ and we have 
$[ {\hat S}_i^i , {\hat B}_r^* ]=0$. 
In the case $i=M-r+1, \cdots, M$, the relation (\ref{B-3}) gives us 
\begin{equation}\label{B-40}
[ {\hat S}_i^i , {\hat B}_r^* ]
=-\left|
\matrix{ {\hat b}_M^{1*} & \cdots & {\hat b}_M^{r*} \cr
         \cdots & \cdots & \cdots \cr
         {\hat b}_{M-r+1}^{1*} & \cdots & {\hat b}_{M-r+1}^{r*} 
         }
         \right|
=-{\hat B}_r^* \ .
\end{equation}
In this case, ${\hat b}_i^{s*}$ is replaced with ${\hat b}_i^{s*}$ 
itself. 
In the case $p=1, 2, \cdots, r$, we have 
\begin{equation}\label{B-50}
[ {\hat T}_p^p , {\hat B}_r^* ]
=\left|
\matrix{ {\hat b}_M^{1*} & \vdots & {\hat b}_M^{r*} \cr
         \vdots & \vdots & \vdots \cr
         {\hat b}_{M-r+1}^{1*} & \vdots & {\hat b}_{M-r+1}^{r*} 
         }
         \right|
={\hat B}_r^* \ .
\end{equation}
In the case $p=r+1, \cdots, M$, ${\hat B}_r^*$ does not 
contain ${\hat b}_k^{p*}$ and we have 
$[ {\hat T}_p^p , {\hat B}_r^* ]=0$. 
By expressing the change of $r$ in terms of $i$ and $p$, 
we obtain the relations (\ref{4-9a}) and (\ref{4-9b}).

Finally, the proof of the relations (\ref{4-14}) is sketched. 
As an example, we will show the case 
$[ {\hat S}^i , {\hat A}_M^* ]=0$. 
The other relations can be proved in the way similar to the case 
$[ {\hat S}^i , {\hat A}_M^* ]=0$. 
For the proof of $[ {\hat S}^i , {\hat A}_M^* ]=0$, 
the following relations must be proved : 
\begin{eqnarray}
& &\left[ \sum_{p=1}^M{\hat a}^{p*}{\hat b}_i^p , {\hat B}_M^* \right]
={\hat B}_M^*(i) \ , \label{B-60}\\
& &\left[ \sum_{p=1}^M{\hat a}^{p*}{\hat b}_i^p , {\hat B}_M^*(l) 
\right]
=0 \ . \label{B-7}
\end{eqnarray}
With the use of the expansion (\ref{B-3}) for $r=M$, we have 
\begin{eqnarray}\label{B-8}
\left[ \sum_{p=1}^M{\hat a}^{p*}{\hat b}_i^p , {\hat B}_M^* \right]
&=&\left[ \sum_{p=1}^M{\hat a}^{p*}{\hat b}_i^p , 
\sum_{q=1}^M{\hat b}_i^{q*} {\hat \Delta}_q^{i*} \right] \nonumber\\
&=&\sum_{q=1}^M{\hat a}^{q*}{\hat \Delta}_q^{i*}
={\hat B}_M^*(i) \ . 
\end{eqnarray}
For the relation (\ref{B-7}), we divide the cases 
$i=l$ and $i\neq l$. 
In the case $i=l$, ${\hat B}_M^*(l)$ does not contain 
${\hat b}_i^{p*}$ and, then, we have the relation (\ref{B-7}). 
In the case $i\neq l$, the commutator becomes of the form that the 
elements 
of the $i$-th row are identical with those of the $l$-th row 
$({\hat a}^{1*}, \cdots, {\hat a}^{M*} )$ and the 
determinant becomes vanish. 
Then, through the following procedure, we get the relation 
$[ {\hat S}^i , {\hat A}_M^* ]=0$ : 
\begin{eqnarray}\label{B-9}
\left[ {\hat S}^i , {\hat A}_M^* \right]
&=& \left[ {\hat a}_i^*{\hat b}+\sum_{p=1}^M{\hat a}^{p*}{\hat b}_i^p , 
{\hat B}_M^*{\hat b}+\sum_{l=1}^M{\hat B}_M^*(l){\hat a}_l \right] 
\nonumber\\
&=&-\sum_{l=1}^M {\hat B}_M^*(l) \left[ {\hat a}_l , {\hat a}_i^* \right] 
{\hat b} 
+\left[ \sum_{p=1}^M{\hat a}^{p*}{\hat b}_i^p , {\hat B}_M^* \right] 
{\hat b}
+\sum_{l=1}^M \left[ \sum_{p=1}^M{\hat a}^{p*}{\hat b}_i^p , 
{\hat B}_M^*(l) \right]{\hat a}_l \nonumber\\
&=&-{\hat B}_M^*(i){\hat b}+{\hat B}_M^*(i){\hat b}
=0 \ .
\end{eqnarray}

\end{document}